\documentclass[sigconf]{acmart}

\usepackage{url}
\usepackage{wrapfig}
\usepackage{algorithm}
\usepackage{algorithmic} 
\usepackage{multirow}
\usepackage{subcaption}

\AtBeginDocument{%
  }

\acmYear{2025}
\setcopyright{cc}
\setcctype{by}
\acmConference[KDD '25]{Proceedings of the 31st ACM SIGKDD Conference on
Knowledge Discovery and Data Mining V.2}{August 3--7, 2025}{Toronto, ON,
Canada}
\acmBooktitle{Proceedings of the 31st ACM SIGKDD Conference on Knowledge
Discovery and Data Mining V.2 (KDD '25), August 3--7, 2025, Toronto, ON,
Canada}
\acmDOI{10.1145/3711896.3736825}
\acmISBN{979-8-4007-1454-2/2025/08}
\begin{document}

\title{A Unified Solution to Diverse Heterogeneities in One-Shot Federated Learning}

\author{Jun Bai}
\authornote{The three authors contributed equally to this research.}
\authornote{Corresponding authors: Jun Bai (jun.bai@mcgill.ca), Yue Li (yueli@cs.mcgill.ca)
}
\affiliation{%
  \institution{McGill University}
  \institution{Mila-Quebec AI Institute}
  \city{Montreal}
  \country{Canada}
}
\email{jun.bai@mcgill.ca}

\author{Yiliao Song}
\authornotemark[1]
\affiliation{%
  \institution{The University of Adelaide}
  \city{Adelaide}
  \country{Australia}}
\email{lia.song@adelaide.edu.au}

\author{Di Wu}
\authornotemark[1]
\affiliation{%
  \institution{University of Southern Queensland}
  \city{Toowoomba}
  \country{Australia}
}
\email{di.wu@unisq.edu.au}

\author{Atul Sajjanhar}
\affiliation{%
 \institution{Deakin University}
 \city{Geelong}
 \country{Australia}}
\email{atul.sajjanhar@deakin.edu.au}

\author{Yong Xiang}
\affiliation{%
 \institution{Deakin University}
 \city{Geelong}
 \country{Australia}}
\email{yong.xiang@deakin.edu.au}

\author{Wei Zhou}
\affiliation{%
  \institution{Monash University}
  \city{Melbourne}
  \country{Australia}}
\email{wei.zhou2@monash.edu}

\author{Xiaohui Tao}
\affiliation{%
  \institution{University of Southern Queensland}
  \city{Springfield}
  \country{Australia}}
\email{xiaohui.tao@unisq.edu.au}

\author{Yan Li}
\affiliation{%
  \institution{University of Southern Queensland}
  \city{Toowoomba}
  \country{Australia}}
\email{yan.li@unisq.edu.au}

\author{Yue Li}
\authornotemark[2]
\affiliation{%
  \institution{McGill University}
  \institution{Mila-Quebec AI Institute}
  \city{Montreal}
  \country{Canada}}
\email{yueli@cs.mcgill.ca}

\renewcommand{\shortauthors}{Jun Bai et al.}

\begin{abstract}
One-Shot Federated Learning (OSFL) restricts communication between the server and clients to a single round, significantly reducing communication costs and minimizing privacy leakage risks compared to traditional Federated Learning (FL), which requires multiple rounds of communication. However, existing OSFL frameworks remain vulnerable to distributional heterogeneity, as they primarily focus on model heterogeneity while neglecting data heterogeneity. To bridge this gap, we propose FedHydra, a unified, data-free, OSFL framework designed to effectively address both model and data heterogeneity. Unlike existing OSFL approaches, FedHydra introduces a novel two-stage learning mechanism. Specifically, it incorporates model stratification and heterogeneity-aware stratified aggregation to mitigate the challenges posed by both model and data heterogeneity. By this design, the data and model heterogeneity issues are simultaneously monitored from different aspects during learning. Consequently, FedHydra can effectively mitigate both issues by minimizing their inherent conflicts. We compared FedHydra with five SOTA baselines on four benchmark datasets. Experimental results show that our method outperforms the previous OSFL methods in both homogeneous and heterogeneous settings. The code is available at https://github.com/Jun-B0518/FedHydra.

\end{abstract}




\maketitle

\section{Introduction}
\label{sec:intro}
Federated Learning (FL) can promise enhanced privacy and data security via collaborative effort in training a global model across multiple clients without sharing private data \cite{mcmahan2017communication}. Existing studies implicate two \textit{limitations} in traditional FL, \textit{i.e.}, \textbf{(I)} it is vulnerable to potential attacks due to its reliance on multi-round communications \cite{10429780, wu2023fedinverse, zhang2019poisoning, jere2020taxonomy}; \textbf{(II)} the performance will be largely impaired under heterogeneity scenarios \cite{mendieta2022local,huang2022learn,karimireddy2020scaffold,acar2020federated}. These issues highlight the need for heterogeneity-tolerant and short-communicated FLs to ensure privacy and efficiency across diverse applications.

One-Shot Federated Learning (OSFL) is proposed to overcome \textit{limitation} \textbf{(I)} by restricting communication to a single round, where all clients are assumed to participate and upload their locally converged models to the server at once \cite{guha2019one}. By this design, OSFL can minimize risks associated with data sharing and communication costs while maintaining data privacy. Meanwhile, existing OSFL shows potential capabilities to address \textit{limitation} \textbf{(II)} \cite{guha2019one, dennis2021heterogeneity, ijcai2021p205, heinbaugh2022data, zhou2020distilled, zhang2022dense}. 
For example, recent methods have used auxiliary public datasets \cite{dennis2021heterogeneity, guha2019one, ijcai2021p205, zhou2020distilled} or embedding client data \cite{heinbaugh2022data} to mitigate the heterogeneity issue.
Despite its potential, OSFL still performs poorly in highly heterogeneous scenarios, which largely prevents OSFL from being applied in reality \cite{mendieta2022local, huang2022learn, qu2022rethinking, tang2022virtual, zhu2021data}. 

\begin{figure}[t]
\centering
\setlength{\abovecaptionskip}{0.2cm}
\includegraphics[width=1\linewidth,scale=1.0]{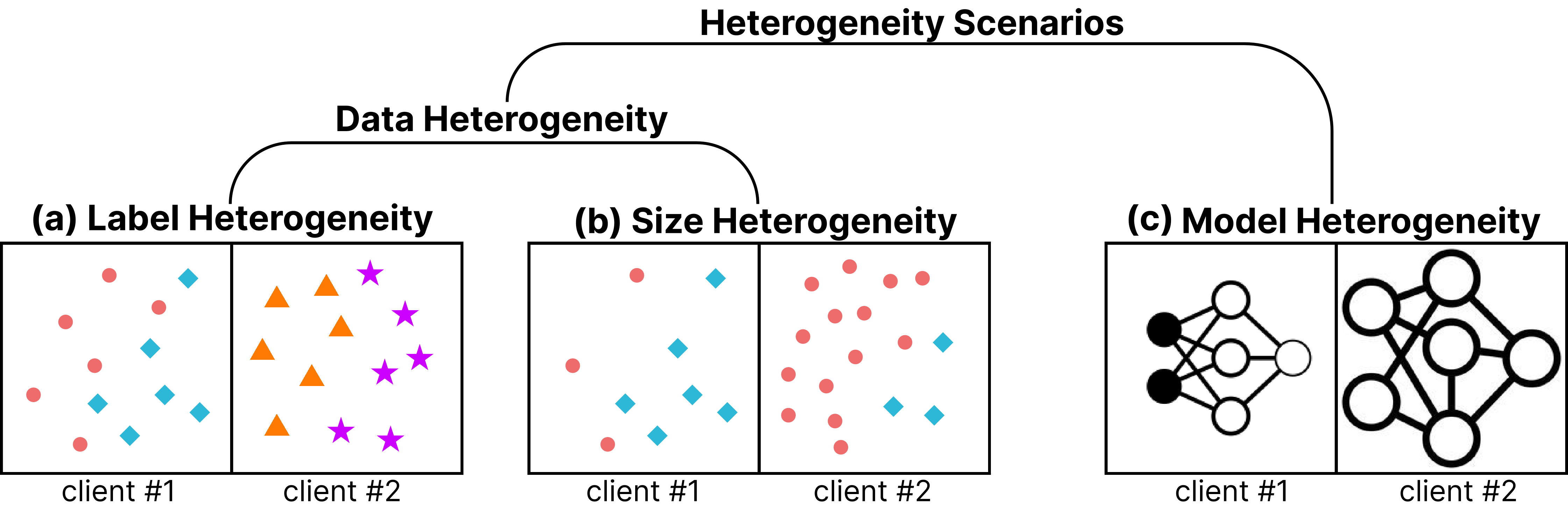}
\caption{Illustration of heterogeneity problems (a) Label Heterogeneity, (b) Size Heterogeneity, and (c) Model Heterogeneity.}
\label{fig:heterogeneity}
\end{figure}

To further discover the cause for low heterogeneity tolerance in existing OSFLs, we review the heterogeneity scenarios from \textit{data} or \textit{model} dimensions. \textit{Data heterogeneity} can be presented as label heterogeneity \textit{i.e.}, Figure~\ref{fig:heterogeneity} (a), which arises when clients have different classes \cite{he2017learning} or size heterogeneity \textit{i.e.}, Figure~\ref{fig:heterogeneity} (b), which refers to the disparity in dataset volumes across clients \cite{he2009learning}. The label heterogeneity can skew global model performance towards more prevalent classes, while size heterogeneity leads to the problem of dominance by larger datasets. \textit{Model heterogeneity} \textit{i.e.}, Figure~\ref{fig:heterogeneity} (c) involves clients employing diverse model architectures \cite{li2020federated}.

Diverse heterogeneity scenarios often coexist, yet existing solutions typically address only one dimension. Methods using auxiliary public datasets mitigate data heterogeneity \cite{dennis2021heterogeneity, guha2019one, ijcai2021p205, zhou2020distilled} but fail to handle model heterogeneity, limiting their applicability to customized models. To eliminate the need for auxiliary datasets and support model heterogeneity, knowledge distillation-based methods like DENSE \cite{zhang2022dense} have been proposed. Co-Boosting \cite{dai2024enhancing}, an extension of DENSE, enhances synthetic data and ensemble models through adversarial sample generation and dynamic client weighting, but still struggles with data heterogeneity. Besides, aggregation-based approaches like OT \cite{singh2020model} facilitate one-shot knowledge transfer by aligning and merging model parameters without retraining. However, data heterogeneity remains a major challenge, weakening performance. These limitations underscore the need for a holistic approach addressing both data and model heterogeneity in OSFL.

Existing methods \cite{zhang2022dense, dennis2021heterogeneity} fail to effectively assess the capability of different clients in handling heterogeneity when aggregating heterogeneous client models on the server. Therefore, they do not adequately address data heterogeneity. Since data heterogeneity originates from clients, its impact first propagates through the client before reaching the server. To mimic this client-server pathway, we propose FedHydra, which operates in two stages. In the \textit{Model Stratification} (\textbf{MS}) stage, FedHydra evaluates each client model's classification capability across all class labels, enabling Stratified Aggregation (\textbf{\textit{SA}}) for heterogeneous client models. Additionally, in the \textit{Heterogeneity-Aware Stratified Aggregation} (\textbf{HASA}) stage, FedHydra further enhances robustness against both data and model heterogeneity through refined \textbf{\textit{SA}}. With this design, FedHydra effectively mitigates data and model heterogeneity issues by minimizing their inherent conflicts, leading to a more stable and efficient OSFL.

The contributions of the proposed FedHydra are as follows:  
1) A \textit{Model Stratification} stage is introduced to assess client models' classification capability across class labels, establishing the foundation for stratified aggregation on heterogeneous client models.  
2) A \textit{Heterogeneity-Aware Stratified Aggregation} stage is designed to tackle both data and model heterogeneity in OSFL through a data-free, stratified aggregation-guided distillation framework.  
3) Extensive evaluations on multiple benchmark datasets under both homogeneous and heterogeneous settings consistently demonstrate FedHydra’s superior efficacy over SOTA baselines.  

\section{Related Works}
\label{sec:related}

FL \cite{mcmahan2017communication}, a decentralized framework, faces challenges from heterogeneity. Strategies to address this include FedAvg \cite{mcmahan2017communication} enhancements with proximal terms \cite{li2020federated, karimireddy2020scaffold, wang2020tackling, acar2020federated, li2021model, singh2020model} and knowledge distillation (KD) to mitigate parameter averaging using auxiliary data \cite{lin2020ensemble, sattler2021fedaux} or generative models \cite{zhu2021data, zhang2022fine}. Recent efforts optimize client selection \cite{tang2022fedcor}, prevent catastrophic forgetting \cite{huang2022learn}, and enhance model adaptability \cite{bai2025non, mendieta2022local}. However, these methods rely on repeated communications, making them unsuitable for OSFL.

OSFL offers efficient learning with minimal communication, employing heuristic client selection and KD with auxiliary datasets for aggregation \cite{guha2019one}. Later research expanded KD to diverse model structures \cite{ijcai2021p205}, but reliance on auxiliary data limits applicability where data sharing is restricted. Secure data transfer methods, including dataset distillation \cite{zhou2020distilled} and data augmentation \cite{shin2020xor}, have been explored. Dense \cite{zhang2022dense} and Co-Boosting \cite{dai2024enhancing} introduced data-free KD using a generator trained on client model ensembles. Heinbaugh et al. \cite{heinbaugh2022data} proposed FEDCVAE-ENS and FEDCVAE-KD for tackling statistical heterogeneity via KD but require local data manipulation, posing data leakage risks. However, many OSFL methods struggle with or remain untested against high statistical heterogeneity, a gap our approach aims to address.

In data-free KD \cite{liu2021data}, DeepInversion \cite{yin2020dreaming} optimizes RGB pixels using cross-entropy and batch normalization losses, while DAFL \cite{chen2019data} employs a generator with the teacher model as a discriminator. ZSKT \cite{micaelli2019zero} generates images to highlight student-teacher discrepancies, reducing overfitting. Unlike methods selecting optimal images for KD, our approach leverages all generated intermediates to mitigate overfitting risks. While Raikward et al. \cite{raikwar2022discovering} use random noise for KD, their method still relies on real images. In contrast, our strategy enables OSFL without real images.

\section{Methodology}
\label{sec:method}

\subsection{FedHydra: An overview}

The proposed FedHydra framework introduces a data-free OSFL training scheme with two key stages: \textbf{MS} and \textbf{HASA}. \textbf{HASA} further comprises two phases: data generation and model distillation. Following the standard OSFL process, all clients transmit their converged local models to the server in a single communication round, where the core implementation of FedHydra takes place. 

\begin{figure*}[t]
\centering
\setlength{\abovecaptionskip}{0.2cm}
\includegraphics[width=0.9\linewidth,scale=1.0]{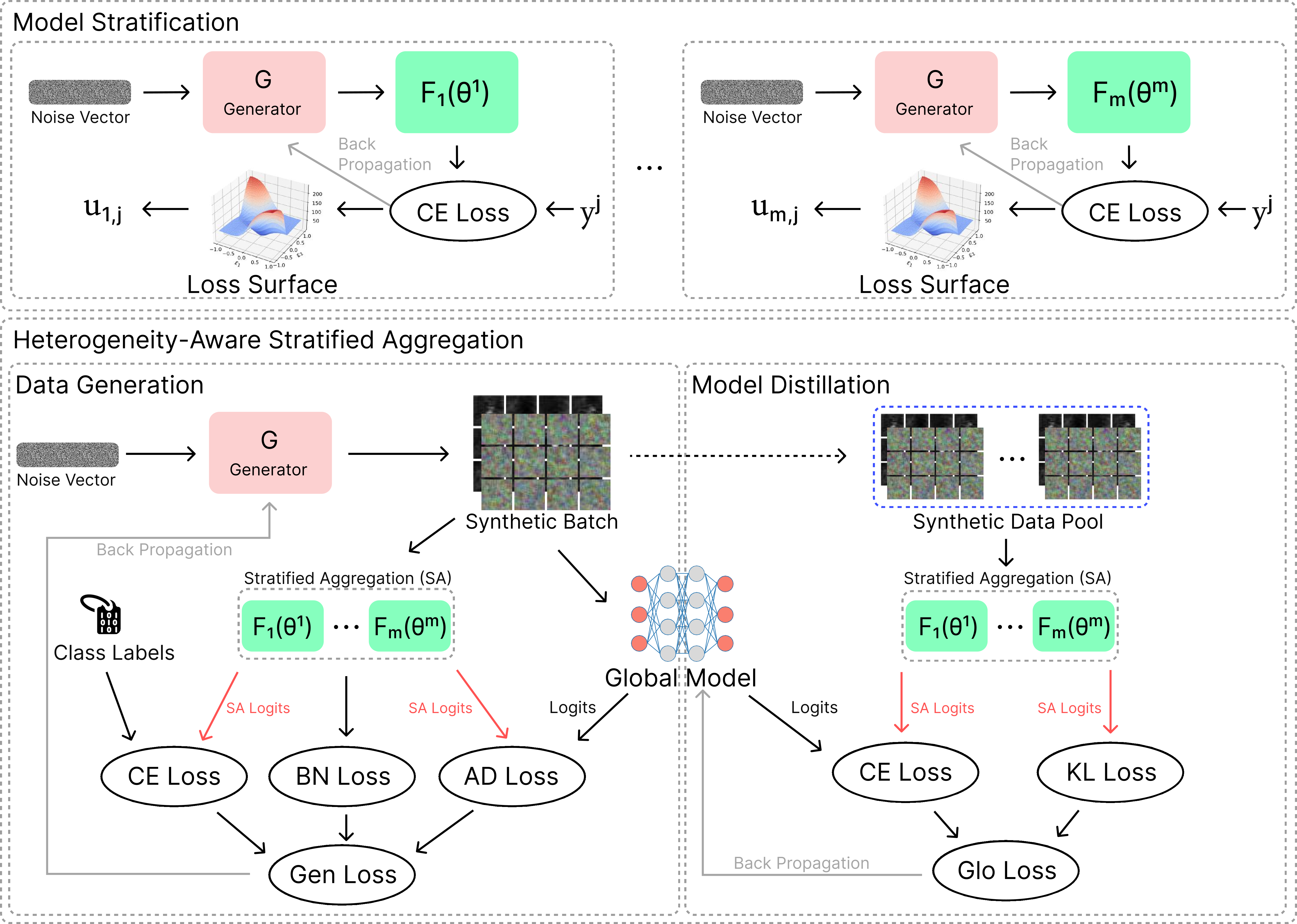}
\caption{Illustration of the FedHydra framework. FedHydra consists of two stages: (1) the \textbf{MS} stage, which evaluates the classification capabilities of uploaded converged client models, and (2) the \textbf{HASA} stage, which enables data-free distillation training of the global model through two phases—data generation and model distillation. }
\label{fig:framework}
\vspace{-0.0cm}
\end{figure*}

Specifically, in \textbf{MS}, we assess client models' classification capabilities across class labels based on generator training loss guided by local models, aiming to refine aggregation weights  These refined aggregation weights inform \textbf{\textit{SA}} for client prediction logits, facilitating both generator and global model training. In \textbf{HASA}, data generation trains a generator using \textbf{\textit{SA}}-based client guidance, producing synthetic data for global model distillation. The two phases alternate, aligning knowledge between synthetic and real data. It is important to note that the designed generator serves different roles in the two stages: it assists in evaluation during \textbf{MS} and generates synthetic data during the aggregation process. Finally, the server constructs a unified global model from client models in a single communication round. The learning process is depicted in Figure~ \ref{fig:framework}, with Algorithm \ref{alg:fedhydra} detailing the full training procedure.

\begin{algorithm}[t]
    \caption{\textit{FedHydra Algorithm}}
    \label{alg:fedhydra}
    \begin{algorithmic}
        \STATE \textbf{Input:} Client models $\{F_k(\cdot)\}_{k=1}^{m}$ with parameters ${\{\theta^k\}_{k=1}^{m}}$, the global model $F_g(\cdot)$ with parameter $\theta^g$, the Generator $G(\cdot)$ with parameter $\theta_G$, class labels $\{y^j\}_{j=1}^{c}$, the global model training epochs $T_g$ and learning rate $\eta_g$, and the generator training epochs $T_G$ and learning rate $\eta_G$
        \STATE \textbf{Output:} Well trained global model $\theta^g$ 
    \end{algorithmic}
    \begin{algorithmic}[1]
        \STATE \textbf{Procedure} \textbf{FedHydra}():
        \STATE \hspace{1em} \textbf{for} each client model $\theta^k$ \textbf{in parallel do}
        \STATE \hspace{2em} $\theta^k \leftarrow$ \textbf{LocalUpdate}($k$)
        \STATE \hspace{2em} Upload converged $\theta^k$ to the server
        \STATE \hspace{1em} \textbf{end for}
        \STATE \hspace{1em} // model stratification
        \STATE \hspace{1em} $\overline{U}_r$, $\overline{U}_c \leftarrow$ \textbf{MS}($\{\theta^k\}_{k=1}^{m}$, $\theta_G$, $\{y^j\}_{j=1}^{c}$)
        \STATE \hspace{1em} \hfill \(\triangleright\) See \textbf{Algorithm \ref{alg:model-interview}}.
        \STATE \hspace{1em} Initialize $\theta^g$ and $\theta_G$
        \STATE \hspace{1em} \textbf{for} each global training epoch $t_g$ from 1 to $T_g$ \textbf{do}
        \STATE \hspace{2em} Sample a batch of noises and labels $\{z_i, y_i\}_{i=1}^b$
        \STATE \hspace{2em} Denoted as $\mathbf{z}=\{z_i\}_{i=1}^{b}$ and $\mathbf{y}=\{y_i\}_{i=1}^{b}$
        \STATE \hspace{2em} // data generation
        \STATE \hspace{2em} \textbf{for} each generator training epoch $t$ from 1 to $T_G$ \textbf{do}
        \STATE \hspace{3em} Generate synthetic data $\hat{\mathbf{x}} = \{\hat{x}_i\}_{i=1}^b \leftarrow G(\mathbf{z}; \theta_G^t)$
        \STATE \hspace{3em} Compute the loss $\mathcal{L}_{Gen}$ by Eq. (\ref{eq:g-loss})
        \STATE \hspace{3em} \hfill \(\triangleright\) See \textbf{Algorithm \ref{alg:cae}} for \textbf{\textit{SA}} computation.
        \STATE \hspace{3em} $\theta_G^{t+1} \leftarrow \theta_G^t - \eta_G \nabla_{\theta_G^t} \mathcal{L}_{Gen}$
        \STATE \hspace{2em} \textbf{end for}
        \STATE \hspace{2em} // model distillation
        \STATE \hspace{2em} Compute the loss $\mathcal{L}_{Glo}$ by Eq. (\ref{eq:md-loss})
        \STATE \hspace{2em} $\theta^g_{t_g+1} \leftarrow \theta^g_{t_g} - \eta_g \nabla_{\theta^g_{t_g}} \mathcal{L}_{Glo}$
        \STATE \hspace{1em} \textbf{end for}
        \STATE \hspace{1em} \textbf{return} $\theta^g$
        \STATE \textbf{End Procedure}
    \end{algorithmic}
\end{algorithm}


\subsection{Model Stratification} 
\label{sec:model-interview}

We first perform \textbf{MS} to assess the classification capabilities of all uploaded converged client models across all class labels using an auxiliary generator model. Specifically, we need to obtain two relative weight matrices. One weight matrix represents the guidance capability of a particular client model across all class labels, and the other represents the guidance capability for a particular class label across all client models. 
 
Assume there are $c$ class labels indexed by $j$ and $m$ clients indexed by $k$ in the federated task. To evaluate the classification capability of one client model for a particular class $j$, we can allow a single client model $F_k(\theta^k)$ to separately guide the training of the Generator $G(\theta_G)$ for generating data with the same class label. In the $t^{th}$ iteration, we can calculate the following cross-entropy (CE) loss:
\begin{equation}
\ell_{k,j}^t(\hat{x}_t^j,y_t^j;\theta_G)=CE(F(\hat{x}_t^j;\theta^k),y_t^j),
\end{equation}
where $\hat{x}_t^j$ and $y_t^j$ denotes the generated samples and assigned labels. After the entire $T_G$ iterations, we obtain the corresponding loss variance denoted as $L_{k,j}=[\ell_{k,j}^1,..,\ell_{k,j}^t,..,\ell_{k,j}^{T_G}]$. Since greater loss variance and  lower minimum loss indicate stronger guidance from the teacher model to the Generator \cite{hu2023toward, cao2021provably, chen2021exploring}, the guidance capability of client model  $F_k(\theta^k)$ to the class $j$ is defined as 
\begin{equation}
    \label{eq:ukj}
  u_{k,j}=\frac{max(L_{k,j})-min(L_{k,j})}{min(L_{k,j})+\epsilon},
\end{equation}
where $\epsilon$ is a small positive constant to prevent division by zero. Based on this, we can obtain a basic guidance capability matrix with a size of $c \times m$ for all client models across all class labels as
\begin{equation}
    U=([(u_{k,j})]_{m \times c})^T.
\end{equation}

Each row of $U$ represents the guidance capability for a specific class label across all client models, while each column reflects the guidance capability of a single client model across all class labels. To realize the \textbf{\textit{SA}} for prediction logits of client models, we need to compare the measure values in $U$ to attain relative importance weights of guidance capability of a particular client model across all class labels as well as for a particular class label across all client models. So we first normalize each row of $U$ to $[0,1]$ by 
\begin{equation}
  \overline{u}_{k,j}=\frac{u_{k,j}}{\sum_{k=1}^{k=m}u_{k,j}}.
\end{equation}

The final normalized $U$ by row can be defined as follows:
\begin{equation}
    \label{eq:Ur bar}
    \overline{U}_r=([(\overline{u}_{k,j})]_{m \times c})^T,
\end{equation}
wherein $\sum_{k=1}^{k=m}\overline{u}_{k,j}=1$. Each row vector of $\overline{U}_r$ represents the stratified weights for prediction logits from all client models for a particular class sample. 

Then, we normalize each column of $U$ to $[0,1]$ as follows:
 \begin{equation}
  \widehat{u}_{k,j}=\frac{u_{k,j}}{\sum_{j=1}^{j=c}u_{k,j}}.
\end{equation}

The final normalized $U$ by column is defined as 
\begin{equation}
    \label{eq:Uc bar}
    \overline{U}_c=([(\widehat{u}_{k,j})]_{m \times c})^T,
\end{equation}
wherein $\sum_{j=1}^{j=c}\widehat{u}_{k,j}=1$. Each column vector of $\overline{U}_c$ represents the stratified weights for prediction logits of a particular client model across all class labels. The detailed procedure for conducting the \textbf{MS} is provided in \textbf{Algorithm \ref{alg:model-interview}}.

\begin{algorithm}[t]
    \caption{\textit{Model Stratification} (\textbf{MS})}
    \label{alg:model-interview}
    \begin{algorithmic}
        \STATE \textbf{Input:} Client models ${\{F_k(\theta^k)\}_{k=1}^{m}}$, the Generator $G(\theta_G)$ and class labels $\{y^j\}_{j=1}^{c}$
        \STATE \textbf{Output:} Stratified weight matrices $\overline{U}_r$ and $\overline{U}_c$
    \end{algorithmic}
    \begin{algorithmic}[1]
        \STATE \textbf{Procedure} \textbf{MS}($\{\theta^k\}_{k=1}^{m}, \theta_G, \{y^j\}_{j=1}^{c}$):
        \STATE \hspace{1em} \textbf{for} each client model $\theta^k$, $k$ from $1$ to $m$ \textbf{do}
        \STATE \hspace{2em} \textbf{for} each class label $y^j$, $j$ from $1$ to $c$ \textbf{do}
        \STATE \hspace{3em} Initialize the Generator parameter $\theta_G$
        \STATE \hspace{3em} Sample a batch of noise vectors and labels
        \STATE \hspace{4em} $\{z_i,y_i^j\}_{i=1}^{b}$, denoted as $\mathbf{z} = \{z_i\}_{i=1}^{b}$ and $\mathbf{y} = \{y_i^j\}_{i=1}^{b}$
        \STATE \hspace{3em} // evaluate the guidance capability by Generator
        \STATE \hspace{3em} \textbf{for} each training epoch $t$ from $1$ to $T_G$ \textbf{do}
        \STATE \hspace{4em} $\hat{\mathbf{x}} = \{\hat{\mathbf{x}}_i\}_{i=1}^{b} \leftarrow G(\mathbf{z}; \theta_G^t)$
        \STATE \hspace{4em} $\ell_{k,j}^t \leftarrow CE(F_k(\hat{\mathbf{x}}; \theta^k), \mathbf{y})$
        \STATE \hspace{4em} $L_{k,j} \leftarrow \ell_{k,j}^t$
        \STATE \hspace{4em} $\theta_G^{t+1} \leftarrow \theta_G^t - \eta_G \nabla_{\theta_G^t} \ell_{k,j}^t$
        \STATE \hspace{3em} \textbf{end for}
        \STATE \hspace{3em} Compute $u_{k,j}$ based on $L_{k,j}$ using Eq. (\ref{eq:ukj})
        \STATE \hspace{3em} $U \leftarrow u_{k,j}$
        \STATE \hspace{2em} \textbf{end for}
        \STATE \hspace{1em} \textbf{end for}
        \STATE \hspace{1em} \parbox[t]{\dimexpr\linewidth-2\algorithmicindent}{%
            Compute $\overline{U}_r$ and $\overline{U}_c$ from $U$ using Eqs. (\ref{eq:Ur bar}) and (\ref{eq:Uc bar})
        }
        \STATE \hspace{1em} \textbf{return} $\overline{U}_r$ and $\overline{U}_c$
        \STATE \textbf{End Procedure}
    \end{algorithmic}
\end{algorithm}

\subsection{Heterogeneity-Aware Stratified Aggregation}

\textbf{HASA} consists of two phases, Data Generation and Model Distillation, both incorporating \textbf{\textit{SA}} to enhance the learning process.

\subsubsection{Stratified Aggregation}
For batch-generated samples by Generator and assigned target labels $\textbf{(x,y)}=\{x_i,y_i^j\}_{i=1}^{b}$ with size $b$, each client model on $\textbf{x}$ will output its batch prediction logits denoted as $P_{k}=F_k(\textbf{x};\theta^{k})$ with size $b \times c$.

Based on Eq. (\ref{eq:Uc bar}), we first conduct an in-model weighted batch prediction logits $\hat{P}_{k}$ for each $P_{k}$ by
\begin{equation}
    \label{eq:pkhar}
    \hat{P}_k = P_k \odot (\mathbf{I}_{b \times 1} \times (\overline{U}_c(:,k))^T),
\end{equation}
wherein $\odot$ means the Hadamard product, $\mathbf{I}_{b \times 1}$ represents a column vector of size $b \times 1$ filled with ones, and $\overline{U}_c(:,k)$ denotes the $k^{th}$ column of $\overline{U}_c$.  Then, we can obtain in-model weighted prediction logits set $P^{i}$ for the $i^{th}$ in-batch sample across all client models as:
\begin{equation}
  \label{eq:p^i}
  P^{i} = \begin{bmatrix}
  \hat{P}_{1}(i,:)   \\
  ..  \\
  \hat{P}_{k}(i,:) \\
  ..  \\
  \hat{P}_{m}(i,:)   \\
  \end{bmatrix}_{m \times c} (i=1,2,..,b), 
\end{equation}
wherein $\hat{P}_{k}(i,:)$ means the $i^{th}$ row of $\hat{P}_{k}$. Subsequently, based on Eq. (\ref{eq:Ur bar}), the weight set for the guidance capability of all client models on batch samples can be calculated by 
\begin{equation}
  \label{eq:V}
  V = \begin{bmatrix}
  \overline{U}_r(\textbf{y}(1),:)   \\
  ..  \\
  \overline{U}_r(\textbf{y}(i),:) \\
  ..  \\
  \overline{U}_r(\textbf{y}(b),:)   \\
  \end{bmatrix}_{b \times m},
\end{equation}
wherein $\textbf{y}(i)$ denotes the $i^{th}$ target class label and $\overline{U}_r(\textbf{y}(i),:)$ means the $\textbf{y}(i)^{th}$ row of $\overline{U}_r$. Combining Eq. (\ref{eq:p^i}) and Eq. (\ref{eq:V}), the inter-model weighted batch prediction logits $P$ can be calculated by
\begin{equation}
  \label{eq:P}
  P = \begin{bmatrix}
   V(1,:) \times P^{1}   \\
  ..  \\
  V(i,:) \times P^{i}  \\
  ..  \\
  V(b,:) \times P^{b}    \\
  \end{bmatrix}_{b \times c},
\end{equation}
wherein $V(i,:)$ represents the $i^{th}$ row of  $V$, and 
$P$ is the final ensemble prediction logits output, termed \textbf{Stratified Aggregation (\textit{SA})}. The specific computation process is summarized in Algorithm \ref{alg:cae}.

\subsubsection{Data Generation}

In this phase, our goal is to train a generator guided by client models to produce synthetic data that closely aligns with the distribution of the aggregated client data. Inspired by \cite{zhang2022dense}, we approach generator training from three key aspects.

We first compute the \textbf{\textit{SA}} prediction logits with Algorithm \ref{alg:cae} for a batch-generated data $\hat{\mathbf{x}}=G(\mathbf{z};\theta_G)$ and their assigned labels $\mathbf{y}$, as:
\begin{equation}
  \label{eq:p-dg}
  P = \textbf{\textit{SA}}(\hat{\mathbf{x}}, \mathbf{y}, \{\theta^k\}_{k=1}^{m}),
\end{equation}
where $P$ is our proposed $\textbf{\textit{SA}}$ logits. Then, to ensure the distribution similarity between synthetic data and real client data, we minimize the cross-entropy (CE) loss between $P$ and $\textbf{y}$, defined in Eq. (\ref{eq:g-ce}), in the training objective function of the generator.
\begin{equation}
\label{eq:g-ce}
\mathcal{L}_{CE}(\hat{\mathbf{x}},\mathbf{y};\boldsymbol{\theta}_G)=CE(P, \mathbf{y}).
\end{equation}

Second, Batch Normalization (BN) loss introduced in \cite{zhang2022dense} is also employed to increase the stability of generator training, defined as:
\begin{equation}
\label{g-abn}
\mathcal{L}_{BN}(\hat{\mathbf{x}};\boldsymbol{\theta}_G)=\frac1m\sum_{k=1}^{m}\sum_l\left(\|\mu_l(\hat{\mathbf{x}})-\mu_{k,l}(\hat{\mathbf{x}})\|+\left\|\sigma_l^2(\hat{\mathbf{x}})-\sigma_{k,l}^2(\hat{\mathbf{x}})\right\|\right),
\end{equation}
where $\mu_{l}(\hat{\mathbf{x}})$ and $\sigma_l^2(\hat{\mathbf{x}})$ are the batch-wise mean and variance of hidden features from the $l^{th}$ BN layer of the generator, $\mu_{k,l}(\hat{\mathbf{x}})$ and $\sigma_{k,l}^{2}(\hat{\mathbf{x}})$ are the mean and variance of the $l^{th}$ BN layer of the $k^{th}$ client model.



\begin{algorithm}[!t]
    \caption{\textit{Stratified Aggregation (\textbf{SA})}}
    \label{alg:cae}
    \begin{algorithmic}
        \STATE \textbf{Input:} Client models ${\{F_k(\theta^k)\}_{k=1}^{m}}$, two weight matrices $\overline{U}_r$ and $\overline{U}_c$ from \textit{Model Stratification}, and synthetic batch data $\mathbf{(\hat{x},y)}=\{x_i,y_i\}_{i=1}^{b}$
        \STATE \textbf{Output:} weighted batch prediction logits matrix $P$
    \end{algorithmic}
    \begin{algorithmic}[1]
        \STATE \textbf{Procedure} \textit{\textbf{SA}}($\{\theta^k\}_{k=1}^{m}, \overline{U}_r, \overline{U}_c, \mathbf{(\hat{x},y)}$):
        \STATE \hspace{1em} \textbf{for} each client model $\theta^k$, $k$ from $1$ to $m$ \textbf{do}
        \STATE \hspace{2em} $P_k \leftarrow F_k(\textbf{x}; \theta^{k})$
        \STATE \hspace{2em} \parbox[t]{\dimexpr\linewidth-\algorithmicindent}{%
        Compute in-model weighted prediction logits $\hat{P}_k$ based on $P_k$ and $\overline{U}_c$ using Eq. (\ref{eq:pkhar})
        }
        \STATE \hspace{1em} \textbf{end for}
        \STATE \hspace{1em} \parbox[t]{\dimexpr\linewidth-\algorithmicindent}{%
    Compute inter-model weighted prediction logits $P$ based on $\{\hat{P}_k\}_{k=1}^{m}$ and $\overline{U}_r$ using Eqs. (\ref{eq:p^i}), (\ref{eq:V}), and (\ref{eq:P})
    }
        \STATE \hspace{1em} \textbf{return} $P$
        \STATE \textbf{End Procedure}
    \end{algorithmic}
\end{algorithm}

Third, to increase the diversity of generated data, an adversarial distillation (AD) loss is considered to maximize the discrepancy between \textbf{\textit{SA}} logits and prediction logits of the global model as:
\begin{equation}
\label{eq:ad}
\mathcal{L}_{AD}(\hat{\mathbf{x}}, \mathbf{y};\boldsymbol{\theta}_G)=- KL\left(P, F_g(\hat{\mathbf{x}};\boldsymbol{\theta}^g)\right),
\end{equation}
where $F_g(\cdot)$ and $\theta^g$ denote the global model and its parameter. Therefore, the final generator training loss is defined by combining the above three loss terms, as follows:
\begin{equation}
\label{eq:g-loss}
\mathcal{L}_{Gen}(\hat{\mathbf{x}},\mathbf{y};\boldsymbol{\theta}_G)=\mathcal{L}_{CE}(\hat{\mathbf{x}},\mathbf{y};\boldsymbol{\theta}_G)+\lambda_1\mathcal{L}_{BN}(\hat{\mathbf{x}};\boldsymbol{\theta}_G)+\lambda_2\mathcal{L}_{AD}(\hat{\mathbf{x}},\mathbf{y};\boldsymbol{\theta}_G),
\end{equation}
where $\lambda_1$ and $\lambda_2$ are adjustable parameters. 

\subsubsection{Model Distillation}

Based on the synthetic data produced in the data generation phase, we design a model distillation scheme to train our global model with knowledge distillation techniques. 

We consider the two loss terms in the training objective function of the global model. First, we use the KL loss between the $\textbf{\textit{SA}}$ logits and prediction logits of the global model to distill the knowledge from client models to the global model, defined as:
\begin{equation}
\label{md-kl}
\mathcal{L}_{KL}(\hat{\mathbf{x}}, \mathbf{y};\boldsymbol{\theta}^g)=KL\left(P,F_g(\hat{\mathbf{x}};\boldsymbol{\theta}^g)\right).
\end{equation}

To achieve efficient and stable model distillation, ensuring consistent prediction labels between the ensemble client models and the global model is crucial, especially considering the potential deviation between synthetic and real data. So, we introduce the following CE loss, defined as:
\begin{equation}
\label{md-ce}
\mathcal{L}_{\widehat{CE}}(\hat{\mathbf{x}}, \mathbf{y};\boldsymbol{\theta}^g)=CE\left(F_g(\hat{\mathbf{x}};\boldsymbol{\theta}^g), H[P]\right),
\end{equation}
where $H[\cdot]$ represents to obtain the hard label of prediction logits. Therefore, by combining the above two loss terms, the final training loss function for realizing model distillation is defined as follows:

\begin{equation}
\label{eq:md-loss}
\mathcal{L}_{Glo}(\hat{\mathbf{x}},\mathbf{y};\boldsymbol{\theta}^g)=\mathcal{L}_{KL}(\hat{\mathbf{x}}, \mathbf{y};\boldsymbol{\theta}^g)+\beta \mathcal{L}_{\widehat{CE}}(\hat{\mathbf{x}}, \mathbf{y};\boldsymbol{\theta}^g),
\end{equation}
where $\beta$ is an adjustable parameter. 

\section{Experiments}

\subsection{Setup}

\subsubsection{Datasets}

Our evaluation experiments utilize four widely used datasets: MNIST \cite{lecun1998gradient}, FashionMNIST \cite{xiao2017fashion}, SVHN \cite{netzer2011reading}, and CIFAR-10 \cite{krizhevsky2009learning}. 
These datasets provide diverse benchmarks for evaluating FL performance.  

\begin{table*}[t]
\setlength{\abovecaptionskip}{1pt}
\fontsize{7.1}{10}\selectfont
\caption{Top-1 test accuracy achieved by baselines and our FedHydra over various datasets with varying $\alpha$.}
\label{table:overview}
\begin{tabular}{c|cccc|cccc|cccc|cccc}
\hline
Dataset         & \multicolumn{4}{c|}{MNIST}                                         & \multicolumn{4}{c|}{FashionMNIST}                                  & \multicolumn{4}{c|}{SVHN}                                          & \multicolumn{4}{c}{CIFAR10}                                       \\ \hline
Method          & $\alpha$=0.5          & $\alpha$=0.3          & $\alpha$=0.1          & $\alpha$=0.01         & $\alpha$=0.5          & $\alpha$=0.3          & $\alpha$=0.1          & $\alpha$=0.01         & $\alpha$=0.5          & $\alpha$=0.3          & $\alpha$=0.1          & $\alpha$=0.01         & $\alpha$=0.5          & $\alpha$=0.3          & $\alpha$=0.1          & $\alpha$=0.01         \\ \hline
FedAvg          & 81.89          & 69.89          & 25.96          & 10.10          & 75.76          & 74.81          & 35.35         & 33.71          & 65.50          & 64.64          & 31.46          & 21.94          & 38.26          & 25.54          & 23.87          & 14.96          \\ \hline
OT          & 77.45          & 77.32          & 50.18          & 25.63          & 74.88          & 66.91          & 41.10          & 39.56          & 78.24          & 69.58          & 55.02          & 15.36          & 52.88          & 47.10          & 36.19          & 30.12          \\ \hline
FedDF           & 78.23          & 73.59          & 31.52          & 10.10          & 73.44          & 65.61        & 46.68  & 40.36                    & 80.20          & 81.00          & 54.35          & 7.59           & 54.81          & 50.82          & 33.06          & 19.91          \\ \hline
DENSE           & 77.86          & 76.78          & 36.73          & 11.17          & 74.20          & 72.79          & 43.26  & 41.65                 & 79.44          & 78.81          & 53.69          & 7.75           & 58.48          & 51.57          & 42.56          & 32.69          \\ \hline
Co-Boosting           & 86.76          & 83.47          & 44.76          & 29.11         & 75.89          & 72.91          & 47.20          & 42.38          & 81.06          & 76.64          & 57.59          & 14.45           & 62.88          & 54.74          & 45.83          & 35.10          \\ \hline
FedHydra (ours) & \textbf{88.75} & \textbf{85.16} & \textbf{59.97} & \textbf{38.83} & \textbf{77.20} & \textbf{76.39} & \textbf{50.48} & \textbf{46.02} & \textbf{82.39} & \textbf{82.96} & \textbf{62.08} & \textbf{30.95} & \textbf{63.15} & \textbf{57.66} & \textbf{48.12} & \textbf{40.63} \\ \hline
\end{tabular}
\end{table*}

\subsubsection{Client Data Distribution} 

\begin{figure}[!ht]
\centering
\setlength{\abovecaptionskip}{0.1cm}
\includegraphics[width=\linewidth,scale=1.0]{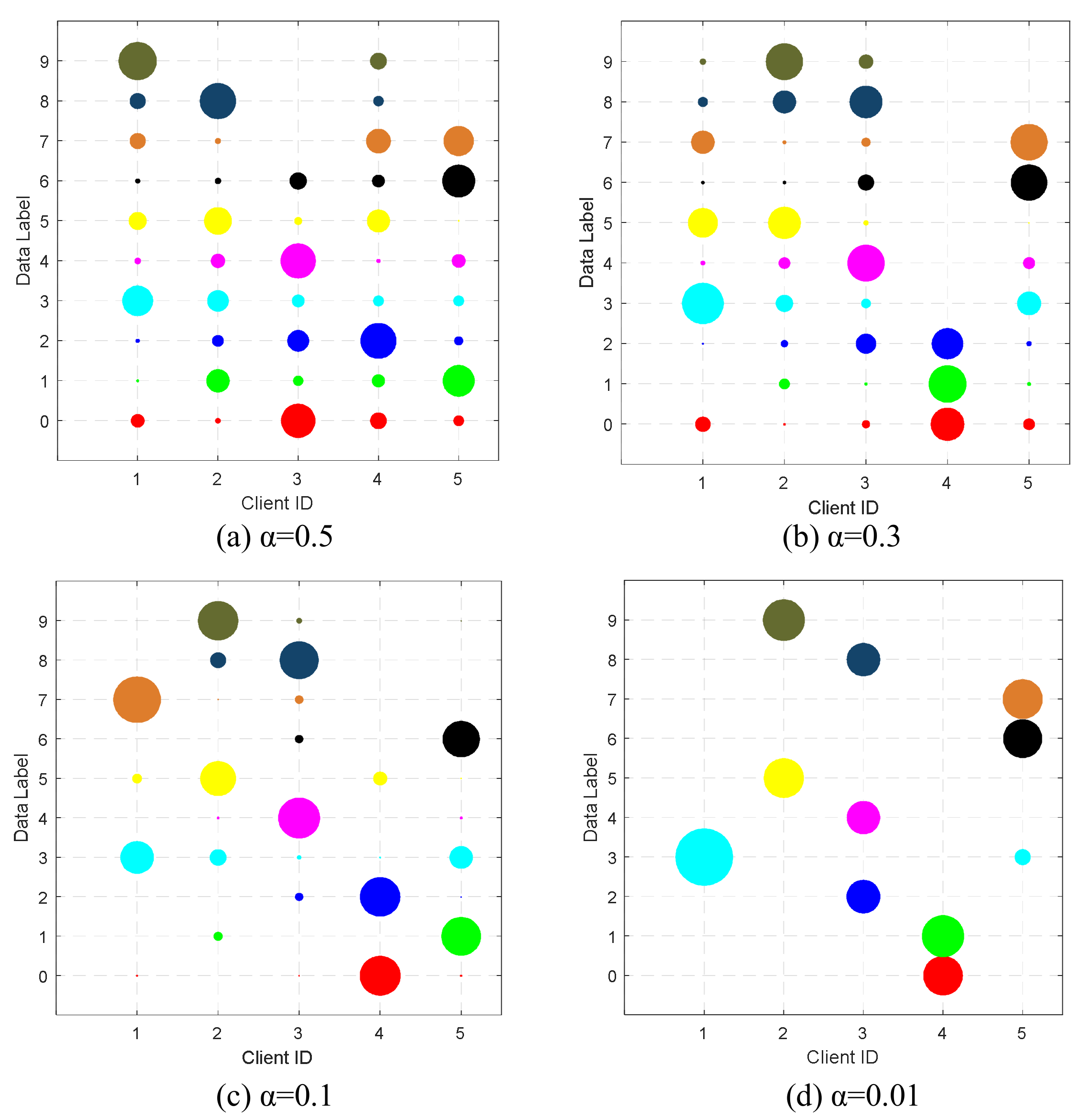}
\caption{Visualization of heterogeneous client data distributions characterized by four diverse $\alpha$ values, with MNIST  as illustration. The horizontal axis denotes the client ID, while the vertical axis represents the class label. Each circle's size corresponds to the number of samples linked to a specific class, and identical-color circles denote the same class label.}
\label{fig:noniid}
\end{figure}

We adopt the Dirichlet distribution method ($Dir(\alpha)$) in \cite{yurochkin2019bayesian} to simulate diverse client data distributions in practical FL scenarios.
Adjusting the parameter $\alpha$  allows us to modify the level of data heterogeneity: a lower $\alpha$ results in more pronounced heterogeneity in the data distribution, while a higher $\alpha$ results in a homogeneous data distribution. We consider four cases, $ \alpha = \{0.5, 0.3, 0.1, 0.01\} $, ranging from a more homogeneous distribution (\(\alpha = 0.5\)) to a highly heterogeneous distribution (\(\alpha = 0.01\)), as illustrated in Figure \ref{fig:noniid}. Unless otherwise specified, we use \( \alpha = 0.5 \) as the default setting.

\subsubsection{Models} The generator introduced in FedHydra begins with a fully connected layer that transforms the input noise vector into a 2D feature map. This is followed by three upsampling convolutional layers, each with batch normalization, LeakyReLU activation, and 2D transposed convolutions. The final layer uses a Sigmoid activation function to output either a $32 \times 32$ RGB image or a $28 \times 28$ grayscale image. For client models, we employ a two-layer CNN for MNIST and FashionMNIST, and a three-layer CNN for SVHN and CIFAR-10 in the case of homogeneous model architecture. In model heterogeneity scenarios, more complex architectures, such as GoogleNet, ResNet18, and LeNet, are utilized in our experiments. 

\subsubsection{Baselines} 

Since the proposed method aims to address the heterogeneity problem in OSFL, we omit the comparison with conventional FL baselines, such as FedProx~\cite{li2020federated}, FedNova~\cite{wang2020tackling}, SCAFFOLD~\cite{karimireddy2020scaffold}, and FedGen~\cite{zhu2021data}, which are specifically designed for multi-round federated optimization rather than single-round FL. Furthermore, we do not compare our method with FEDCVAE~\cite{heinbaugh2022data}, as it directly embeds the data on the client side and uploads it to the server without client-side model training. 
Instead, we evaluate the performance of FedHydra against five SOTA baselines: FedAvg~\cite{mcmahan2017communication}, OT~\cite{singh2020model}, FedDF~\cite{lin2020ensemble}, DENSE~\cite{zhang2022dense}, and Co-Boosting~\cite{dai2024enhancing}.

\subsubsection{Hyperparameter Configurations} 

In our OSFL setup, we consider various numbers of clients $m=\{5, 10, 20, 50, 100\}$, default $m=5$. For the client model training, we employ the SGD optimizer with learning rate $\eta=0.01$, while setting the local epochs and batch size as $B=128$ and $E=200$, respectively. On the server side, we utilize the Adam optimizer with learning rate $\eta_G=0.001$ to train the generator. This generator, characterized by adjustable parameters $\lambda_1=1.0, \lambda_2=1.0$, undergoes training for $T_G=30$ rounds to generate batch examples. For the distillation training of the global model, we set the training epochs to $T_g=200$ and the learning rate to $\eta_g=0.01$, with SGD serving as the optimizer.

\subsection{Evaluation}

\subsubsection{Performance Overview}

FedHydra demonstrates superior performance across various data heterogeneity settings compared to existing methods, including FedAvg, OT, FedDF, DENSE, and Co-Boosting. As shown in Table \ref{table:overview}, FedHydra consistently achieves the highest accuracy across all datasets and values of $\alpha$, indicating strong robustness against heterogeneous data distribution.

\begin{figure}[]
\centering
\setlength{\abovecaptionskip}{0.1cm}
\includegraphics[width=\linewidth,scale=1.0]{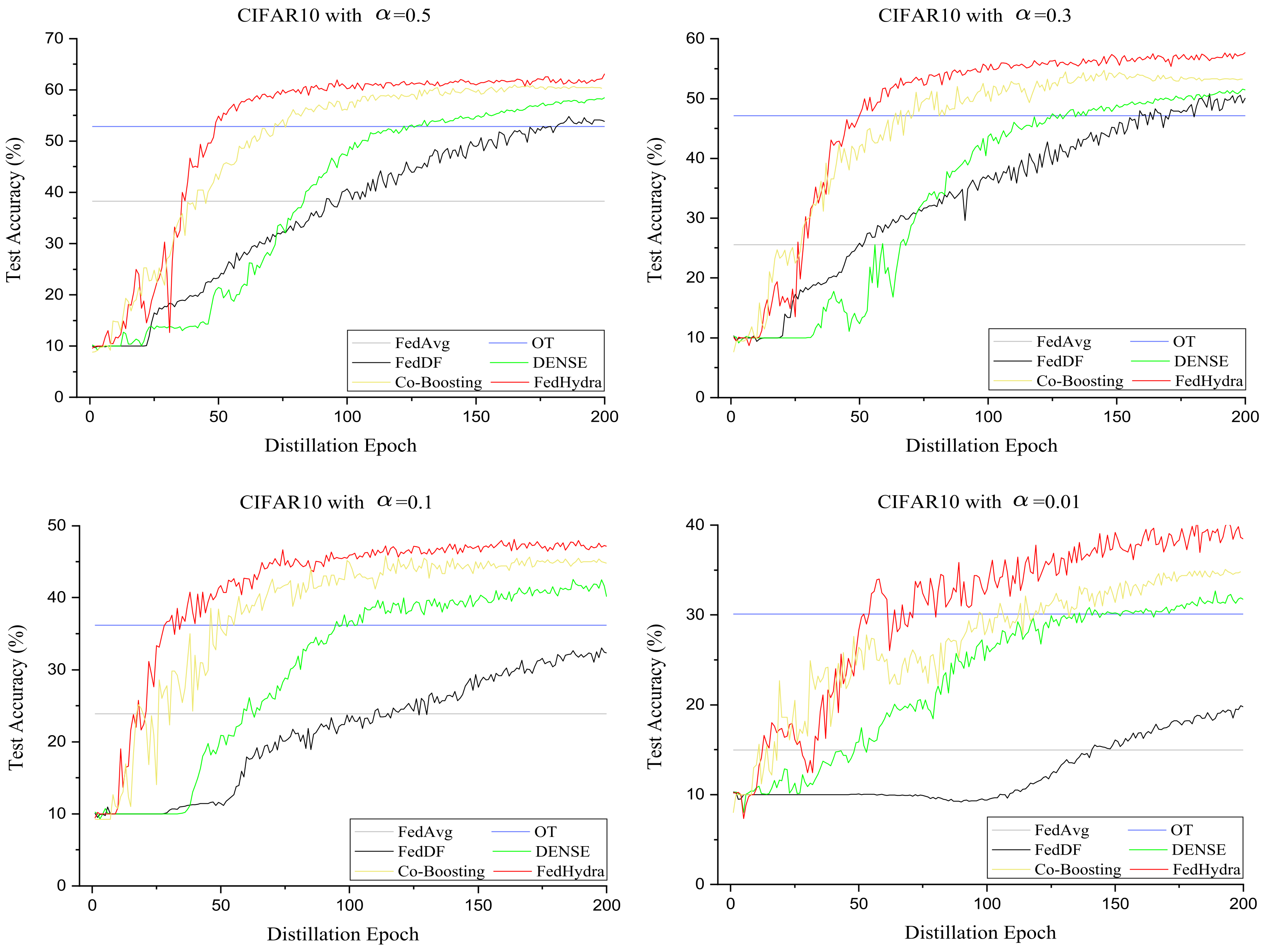}
\caption{Test accuracy versus distillation epoch over CIFAR10 with $\alpha=\{0.5,0.3,0.1,0.01\}$.}
\label{fig:po-cifar10-acc}
\end{figure}

For MNIST, most methods perform well under low heterogeneity ($\alpha = 0.5$) but degrade as it increases. FedAvg drops sharply from 81.89\% to 10.10\%, with FedDF and DENSE showing similar trends. Co-Boosting improves stability but is outperformed by FedHydra, which achieves 38.83\% at $\alpha = 0.01$, surpassing Co-Boosting by 9.72\%. On FashionMNIST and SVHN, FedHydra maintains the highest accuracy, outperforming the second-best method by up to 9.01\% at $\alpha = 0.01$. For CIFAR10, FedHydra reaches 40.63\%, exceeding Co-Boosting by 5.53\% and significantly outperforming FedAvg and OT under extreme data heterogeneity.
Figure~\ref{fig:po-cifar10-acc} presents the training accuracy curves over distillation epochs on CIFAR-10 under $\alpha=\{0.5, 0.3, 0.1, 0.01\}$. Additional curves for MNIST, FashionMNIST, and SVHN are provided in Appendix~\ref{appendix_sec:po}.

FedHydra’s robustness comes from its superior handling of severe data heterogeneity. While FedAvg struggles with parameter averaging and both FedDF and DENSE fail under extreme settings, Co-Boosting improves but remains inferior. These results confirm FedHydra as the most effective OSFL method in heterogeneous environments. Further experiments analyze accuracy trends over distillation epochs across four datasets. 

\begin{figure}[]
\centering
\setlength{\abovecaptionskip}{0.1cm}
\includegraphics[width=\linewidth,scale=1.0]{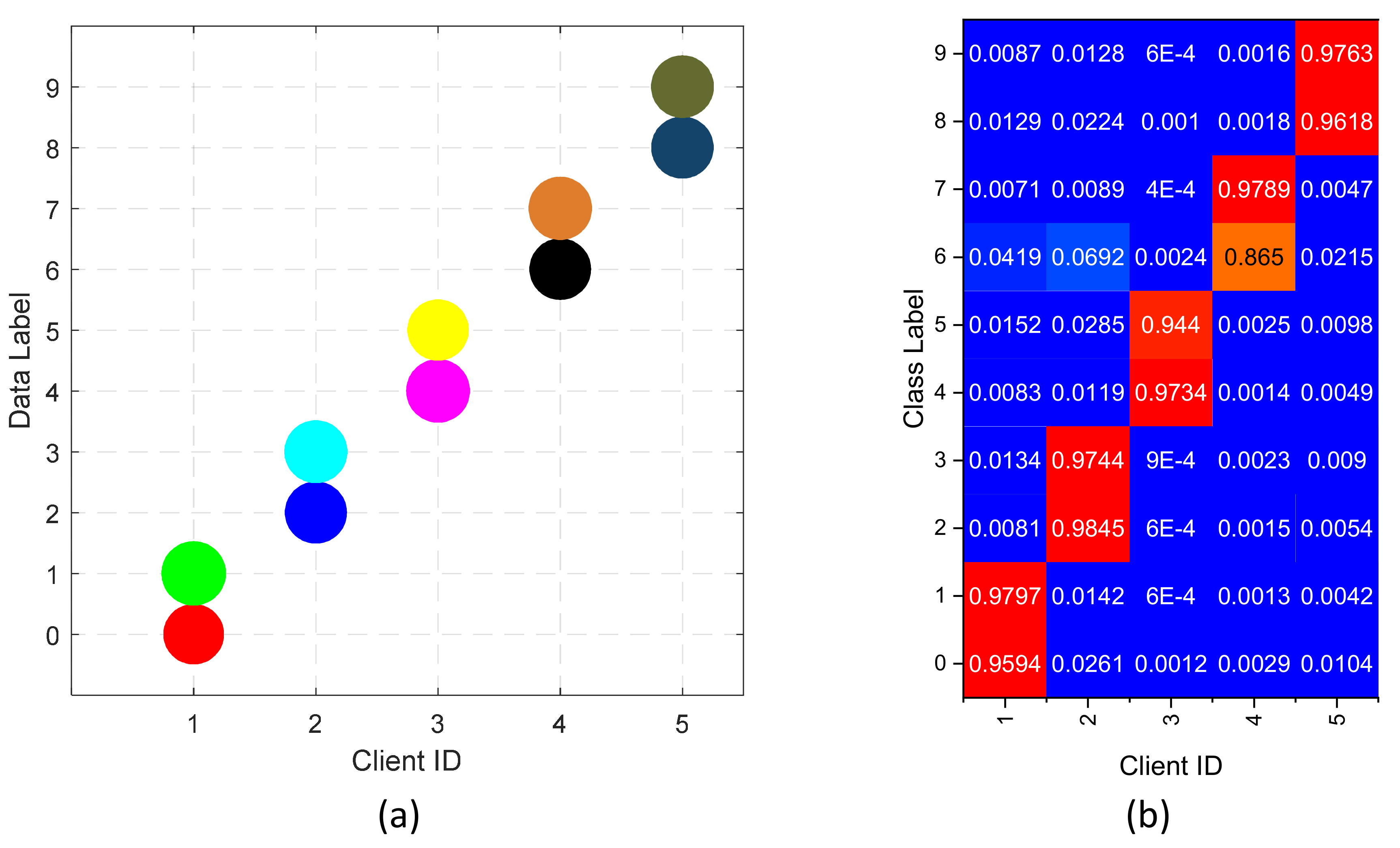}
\caption{Visualization of \textbf{\textit{SA}}. (a) 2c/c data distribution across five clients using MNIST as an example. (b) The weight distribution of evaluated classification capability for five client models on ten class labels.}
\label{fig:2cc-weights}
\vspace{-0.4cm}
\end{figure}

\begin{figure}[b]
\centering
\setlength{\abovecaptionskip}{0.1cm}
\includegraphics[width=\linewidth,scale=1.0]{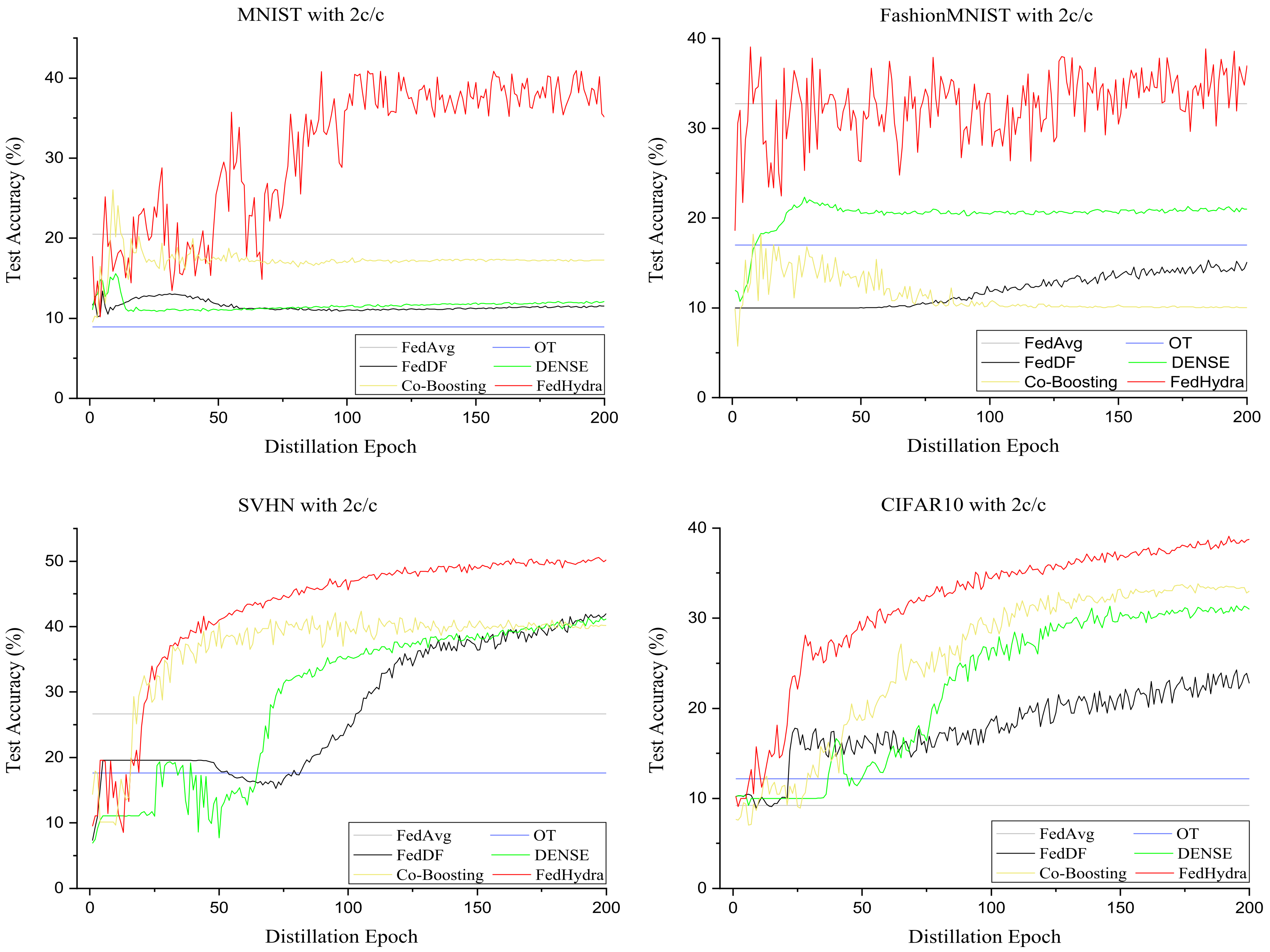}
\caption{Test accuracy versus distillation epoch over diverse datasets with 2c/c distribution.}
\label{fig:2cc-acc}
\end{figure}

\subsubsection{Impact of Stratified Aggregation}

To validate the effectiveness of \textbf{\textit{SA}} under heterogeneous federated distributions, we introduce an extreme \textbf{2c/c} scenario where each client possesses only two disjoint class samples. Figure~\ref{fig:2cc-weights} (a) illustrates this distribution for MNIST across five clients. As shown in Figure~\ref{fig:2cc-weights} (b), each client model excels primarily in its two assigned classes, with classification weights near 1 (e.g., client 1 has $0.9594$ for class 0 and $0.9797$ for class 1, while other weights are close to zero). This confirms that our \textbf{MS} accurately captures client classification capabilities.

To evaluate the performance under \textbf{2c/c} distribution, Table \ref{table:cae-2cc} presents the top-1 test accuracy on four datasets. FedHydra consistently outperforms all baseline methods, achieving the highest accuracy. Among the baselines, OT performs the worst across all datasets. Co-Boosting shows competitive results, especially on SVHN and CIFAR10, but still lags behind FedHydra. FedDF and DENSE perform worse than FedAvg on simpler datasets (MNIST, FashionMNIST), while FedAvg struggles on complex datasets (SVHN, CIFAR10). These results confirm the robustness of FedHydra under \textbf{2c/c} distribution. Test accuracy trends are shown in Figure~\ref{fig:2cc-acc}, further demonstrating FedHydra’s superiority.

\begin{figure}[]
\centering
\setlength{\abovecaptionskip}{0.1cm}
\includegraphics[width=\linewidth,scale=1.0]{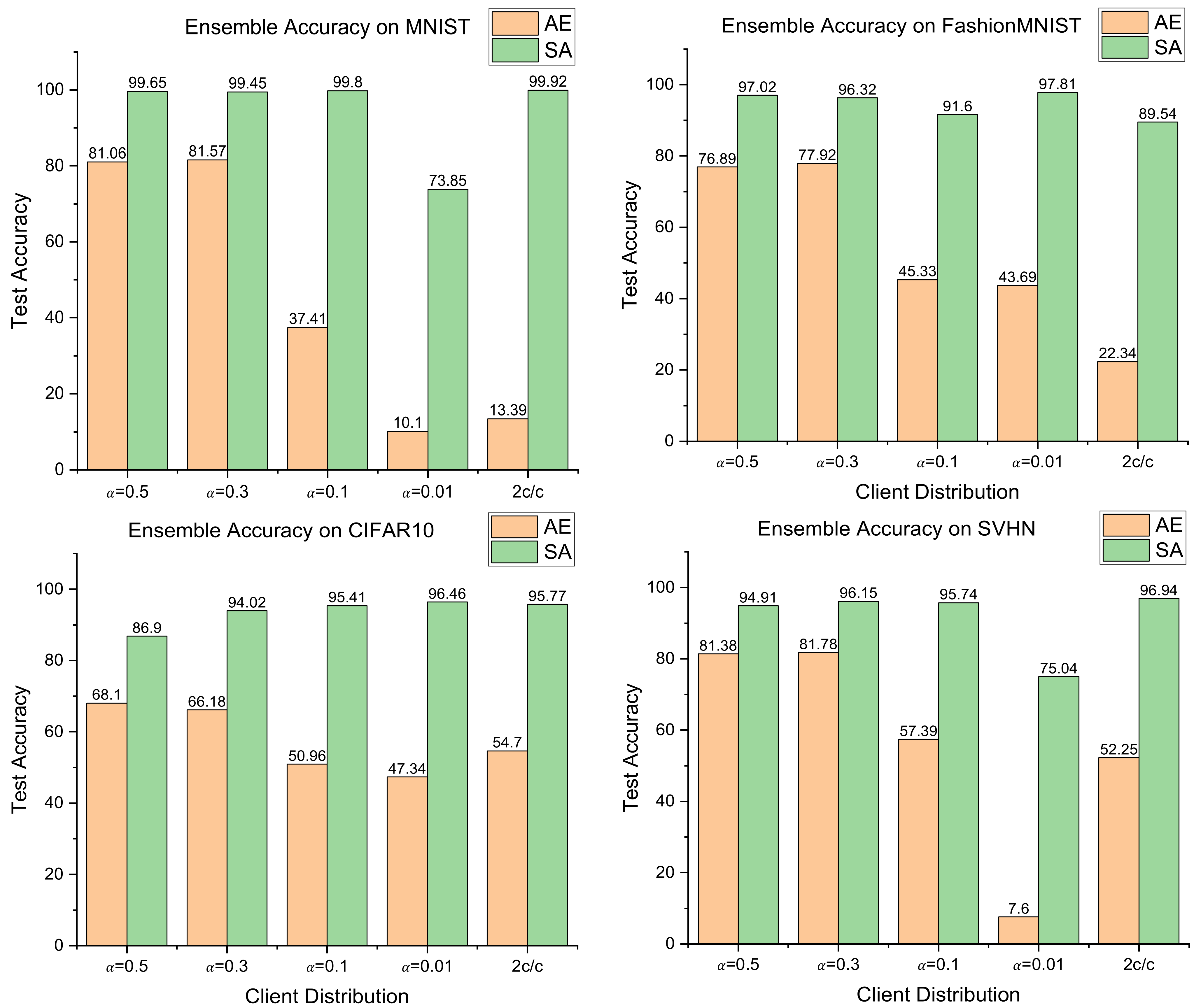}
\caption{Test accuracy achieved by \textbf{\textit{AE}} and our \textbf{\textit{SA}} over four datasets with varying client data distributions.}
\label{fig:cae-vs-ae}
\end{figure}

\begin{table}[]
\setlength{\abovecaptionskip}{5pt}
\fontsize{8}{10}\selectfont
\caption{Top-1 test accuracy achieved by baselines and FedHydra over various datasets with 2c/c distribution.}
\label{table:cae-2cc}
\begin{tabular}{c|cclc}
\hline
Dataset         & MNIST          & FashionMNIST   & SVHN           & CIFAR10        \\ \hline
FedAvg          & 20.50          & 32.73          & 26.68          & 9.22           \\ \hline
OT          & 8.92          & 16.98          & 17.63          & 12.15           \\ \hline
FedDF           & 13.38          & 15.33          & 42.03          & 24.28          \\ \hline
DENSE           & 15.58          & 22.31          & 41.38          & 31.36          \\ \hline
Co-Boosting           & 26.05          & 18.22          & 42.38          & 33.84          \\ \hline
FedHydra (ours) & \textbf{40.81} & \textbf{39.06} & \textbf{50.61} & \textbf{39.09} \\ \hline
\end{tabular}
\end{table}

Furthermore, we compare the test accuracy of the averaging ensemble (\textbf{\textit{AE}}) from DENSE and our \textbf{\textit{SA}} across four datasets, as shown in Figure~\ref{fig:cae-vs-ae}. The results show that \textbf{\textit{SA}} consistently outperforms \textbf{\textit{AE}}, with up to a $20.00\%$ gap at $\alpha = 0.5$ and $0.3$. As data heterogeneity increases, \textbf{\textit{AE}} performance drops sharply, e.g., from $81.75\%$ to $37.41\%$ on MNIST when $\alpha = 0.1$. The gap is largest at $\alpha = 0.01$, which represents both label and size heterogeneity, whereas \textbf{2c/c} maintains equal sizes.

\subsubsection{Impact of Stratified Aggregation Guided Model Distillation}

\begin{figure}[]
\centering
\setlength{\abovecaptionskip}{0.1cm}
\includegraphics[width=\linewidth,scale=1.0]{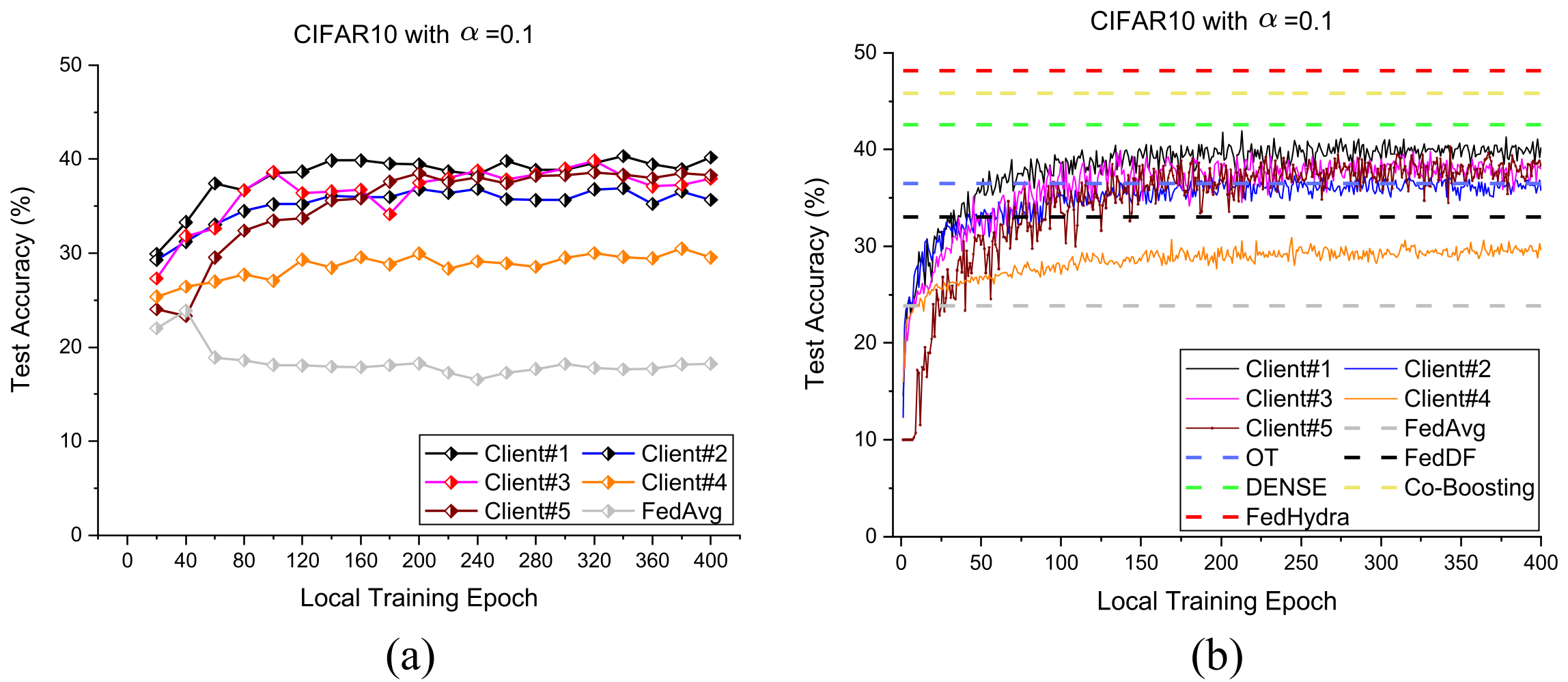}
\caption{(a) Test accuracy of FedAvg and client models across local training epochs $E={20,40,60,...,400}$. (b) Local training accuracy of different clients. The dotted lines represent the top-1 test accuracy achieved by baselines and our FedHydra.}
\label{fig:impact-md-cifar10-0.1}
\end{figure}

We evaluate the impact of our \textbf{\textit{SA}}-guided model distillation by comparing it with the five SOTA methods, and FedHydra on CIFAR10 with $\alpha=0.1$. The results are illustrated in Figure~\ref{fig:impact-md-cifar10-0.1}. Figure~\ref{fig:impact-md-cifar10-0.1} (a) shows test accuracy across local training epochs $E = \{40, 80, 120, ..., 400\}$. The FedAvg achieves its best accuracy ($\sim24.00\%$) at $E=40$, but performance deteriorates with increasing $E$ due to weight divergence under data heterogeneity. Figure~\ref{fig:impact-md-cifar10-0.1} (b) compares different methods over $E \in [0, 400]$. FedHydra consistently outperforms SOTA methods, and client models, while FedAvg performs worse than individual clients. OT and Co-Boosting improve over FedAvg but remain below FedHydra. These results confirm the robustness of \textbf{\textit{SA}}-guided model distillation, demonstrating its superior handling of heterogeneous data distributions.

\begin{table}[]
\setlength{\abovecaptionskip}{5pt}
\setlength{\tabcolsep}{1pt}
\fontsize{7.5}{10}\selectfont
\caption{Top-1 test accuracy achieved by baselines and FedHydra over CIFAR10 with personalized client models. M1 to M5 correspond to GoogleNet, ResNet18, two customized CNN models, and LeNet, respectively.}
\label{table:hfl-cifar10}
\begin{tabular}{c|ccccc|cccc}
\hline
\multirow{2}{*}{\begin{tabular}[c]{@{}c@{}} $\alpha$ \end{tabular}} & \multicolumn{5}{c|}{Personalized Client Model}                                        & \multicolumn{4}{c}{Server (ResNet18)} \\ \cline{2-10} 
                                                                                      & \multicolumn{1}{c}{M1} & \multicolumn{1}{c}{M2} & M3   & M4   & M5  & FedDF        & DENSE   & Co-Boosting     & FedHydra              \\ \hline
$\alpha$=0.01                                                                                & 26.68                         & 28.78                        & 26.71 & 19.28 & 10.00 & 22.25        & 18.67    & 25.63    & \textbf{30.51}        \\ \hline
$\alpha$=0.1                                                                                 & 37.36                         & 36.63                        & 35.60 & 30.41 & 28.77 & 27.39        & 26.43  & 26.82        & \textbf{30.90}        \\ \hline
$\alpha$=0.3                                                                                 & 46.41                         & 55.49                        & 45.78 & 47.98 & 33.47 & 46.96        & 44.54  & 45.61        & \textbf{48.80}        \\ \hline
$\alpha$=0.5                                                                                 & 49.33                         & 61.08                        & 47.65 & 54.14 & 48.25 & 56.27        & 51.29   & 53.76      & \textbf{59.23}        \\ \hline
\end{tabular}

\end{table}

\subsubsection{Performance Comparison under OSFL Model Heterogeneity}

FedHydra effectively addresses data and model heterogeneity. We evaluate its performance on CIFAR10 using five client models (GoogleNet, ResNet18, two CNNs, and LeNet) across varying $\alpha$ values. Since FedAvg and OT do not support model heterogeneity, we compare FedHydra with FedDF, DENSE, and Co-Boosting. As shown in Table \ref{table:hfl-cifar10}, FedHydra consistently outperforms all baselines, achieving $59.23\%$ at $\alpha=0.5$. As $\alpha$ decreases, FedDF and DENSE drop sharply to $22.25\%$ and $18.67\%$ at $\alpha=0.01$, while FedHydra maintains robustness with $30.51\%$, demonstrating superior adaptability. Additional results on SVHN are in Appendix \ref{sec:D}.

\subsubsection{Impact of Client Numbers}

\begin{table}[!h]
\setlength{\abovecaptionskip}{5pt}
\setlength{\tabcolsep}{2pt} 
\fontsize{8}{10}\selectfont
\caption{Top-1 test accuracy achieved by baselines and FedHydra over SVHN with different client numbers $K=\{5,10,20,50,100\}$.}
\label{table:impact-cn-svhn}
\begin{tabular}{c|cccccc}
\hline
Client $K$ & FedAvg  & OT  & FedDF  & DENSE & Co-Boosting & FedHydra  \\ \hline
5   & 65.50 & 78.24 & 80.20  & 79.44 & 81.06 & \textbf{82.39}  \\ \hline
10  & 43.45  & 58.62 & 70.20  & 69.51 & 75.92 & \textbf{77.13}  \\ \hline
20  & 47.87  & 56.76 & 69.36  & 69.10 & 70.01 & \textbf{76.54}  \\ \hline
50  & 36.77 & 45.68 & 68.30  & 67.74 & 67.93 & \textbf{74.89}  \\ \hline
100 & 23.69  & 35.23 & 50.37  & 48.87 & 53.64  & \textbf{60.01}  \\ \hline
\end{tabular}
\end{table}

The number of clients ($K$) significantly impacts OSFL performance, as increasing $K$ intensifies data fragmentation and heterogeneity. Table \ref{table:impact-cn-svhn} shows that as $K$ grows, accuracy declines across all methods. FedHydra consistently achieves the highest accuracy. FedDF and DENSE remain stable but drop significantly at $K=100$, while Co-Boosting struggles with larger $K$. FedAvg and OT perform the worst, with FedAvg falling from $65.50\%$ at $K=5$ to $23.69\%$ at $K=100$. These results highlight FedHydra’s robustness in large-client federated settings. More CIFAR10 experiments are in Appendix \ref{sec:C.1}.

\subsubsection{Performance Under Multiple Global Training Rounds}

\begin{table}[!h]
\setlength{\abovecaptionskip}{5pt}
\setlength{\tabcolsep}{2pt} 
\fontsize{8}{10}\selectfont
\caption{Top-1 test accuracy achieved by baselines and FedHydra over CIFAR10 with $\alpha=0.1$ under multiple global rounds.}
\label{table:mr-cifar10-a0.1}
\begin{tabular}{c|cccccc}
\hline
\begin{tabular}[c]{@{}c@{}}Global \\ round\end{tabular} & FedAvg  & OT  & FedDF  & DENSE & Co-Boosting & FedHydra  \\ \hline
$T$=1  & 22.72 & 18.33 & 21.33 & 29.60 & 31.68 & \textbf{37.95} \\ \hline
$T$=2  & 37.32 & 26.38 & 28.63 & 32.39 & 37.93 & \textbf{46.52} \\ \hline
$T$=3  & 35.95 & 39.54 & 30.36 & 36.42 & 46.23 & \textbf{48.75} \\ \hline
$T$=4  & 43.86 & 36.88 & 30.71 & 38.85 & 46.84 & \textbf{50.17} \\ \hline
$T$=5  & 41.99 & 39.96 & 28.82 & 37.27 & 45.20 & \textbf{49.34} \\ \hline
\end{tabular}
\end{table}

\begin{table}[!h]
\setlength{\abovecaptionskip}{5pt}
\setlength{\tabcolsep}{2pt} 
\fontsize{8}{10}\selectfont
\caption{Top-1 test accuracy achieved by FedAvg, OT, FedDF, DENSE, Co-Boosting,  and FedHydra over CIFAR10 with 2c/c under multiple global rounds.}
\label{table:mr-cifar10-2cc}
\begin{tabular}{c|cccccc}
\hline
\begin{tabular}[c]{@{}c@{}}Global \\ round\end{tabular} & FedAvg  & OT  & FedDF  & DENSE & Co-Boosting & FedHydra  \\ \hline
$T$=1  & 12.23 & 13.86 & 17.84 & 23.57 & 27.78 & \textbf{28.34} \\ \hline
$T$=2  & 15.62 & 15.33 & 18.61 & 22.05 & 22.66 & \textbf{25.54} \\ \hline
$T$=3  & 16.10 & 18.66 & 16.76 & 25.97 & 29.00 & \textbf{32.25} \\ \hline
$T$=4  & 20.44 & 22.36 & 15.81 & 20.06 & 34.11 & \textbf{35.69} \\ \hline
$T$=5  & 18.24 & 21.02 & 17.28 & 24.24 & 34.88 & \textbf{38.76} \\ \hline
\end{tabular}
\end{table}


Global training rounds ($T$) play a vital role in FL performance. Table~\ref{table:mr-cifar10-a0.1} and Table~\ref{table:mr-cifar10-2cc} report CIFAR10 accuracy at $\alpha=0.1$ and under the 2c/c setting, respectively. In both cases, FedHydra consistently outperforms all baselines across different $T$ values. It achieves 50.17\% at $T=4$ on $\alpha=0.1$ and 38.76\% at $T=5$ under 2c/c, maintaining a clear lead. In contrast, FedAvg and OT fluctuate, while FedDF often underperforms. Co-Boosting and DENSE improve with more rounds but remain below FedHydra. These results confirm FedHydra's robustness and efficiency in multi-round FL, even under heterogeneous conditions. More SVHN results are provided in Appendix~\ref{sec:C.2} and~\ref{sec:E}.

\begin{figure}[]
\centering
\setlength{\abovecaptionskip}{0.3cm}
\includegraphics[width=0.8\linewidth,scale=0.8]{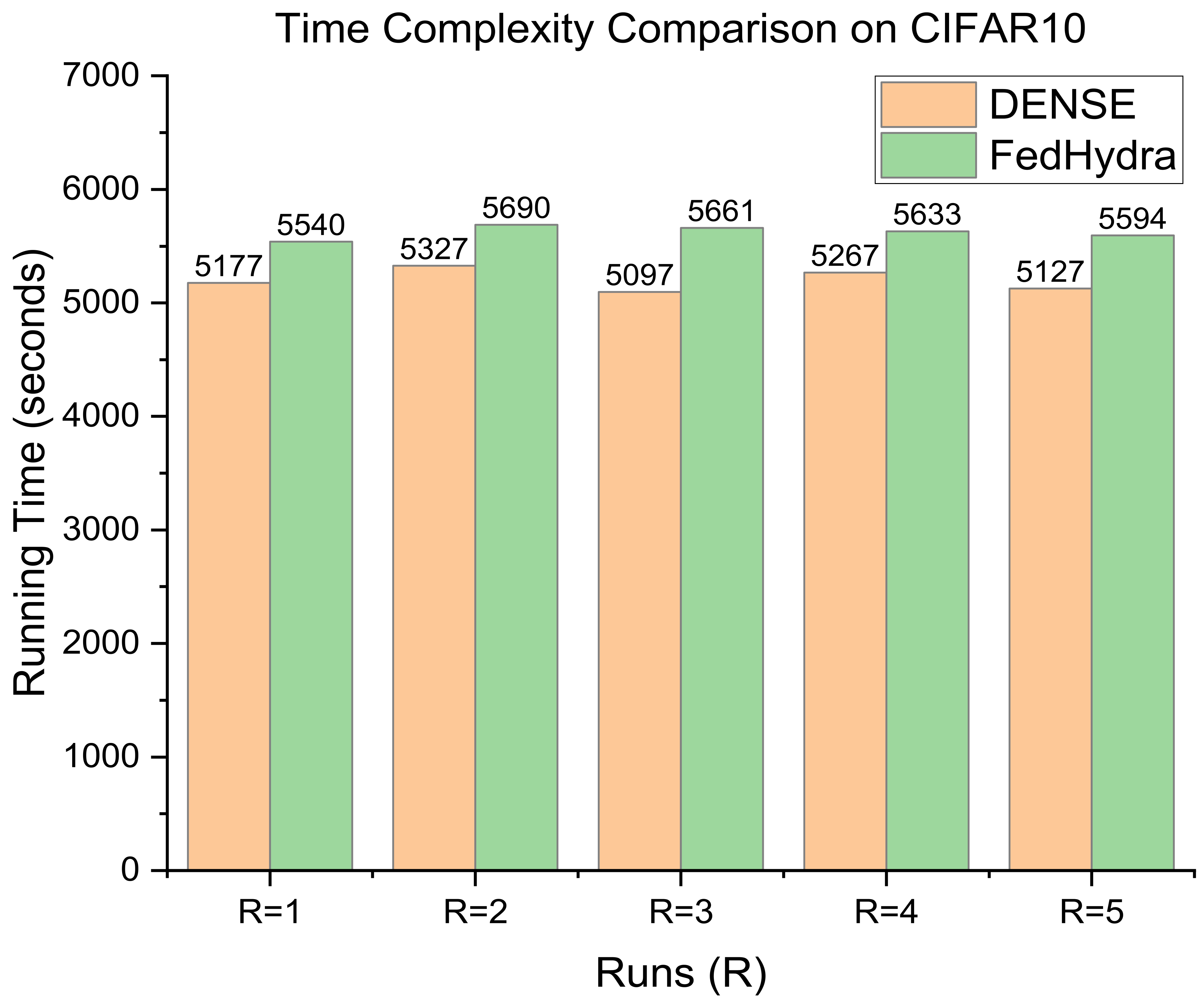}
\caption{Time complexity comparison of FedHydra and DENSE based on five repeated runs (seconds) on CIFAR10.}
\label{fig:tc-cifar10}
\vspace{-0.1cm}
\end{figure}

\subsubsection{Time Complexity Analysis}

We briefly compare the time complexity of our FedHydra algorithm with DENSE, the primary baseline. Assume that the time complexity (\textit{\textbf{TC}}) of completing one batch training round of generator is $O(1)$. Without considering the \textit{\textbf{TC}} of the global model training, the total \textit{\textbf{TC}} for DENSE is $O(T_g \cdot T_G)$ while for our FedHydra it is $O((m \cdot c + T_g) \cdot T_G)$. Considering the computation capability of the central server and limited client numbers $m$ and classes $c$, we believe the increased \textit{\textbf{TC}} of $O(m \cdot c \cdot T_G)$ is acceptable in practical FL applications when comparing between FedHydra and DENSE. The results of five repeated experiments with running time (seconds) on CIFAR10 are shown in Figure~\ref{fig:tc-cifar10}, where we compare the computational complexity of FedHydra and DENSE. Our results show that FedHydra performs better in heterogeneous scenarios, with only a roughly 7\% increase in computational complexity compared to DENSE, considered acceptable.

\subsubsection{Ablation Study}

\begin{table}[]
\setlength{\abovecaptionskip}{5pt}
\setlength{\tabcolsep}{5pt} 
\fontsize{8}{10}\selectfont
\caption{Top-1 test accuracy achieved by FedHydra over various datasets with varying $\lambda_1$ and $\lambda_2$.}
\label{table:lamda}
\centering
\begin{tabular}{cccccc}
\hline
\textbf{$\lambda_1$} & \textbf{$\lambda_2$} & MNIST & \multicolumn{1}{l}{FashionMNIST} & \multicolumn{1}{l}{SVHN} & \multicolumn{1}{l}{CIFAR10} \\ \hline
1.0             & 1.0             & 88.75          & 77.20                                     & 82.39                             & 63.15                                \\ \hline
0.5             & 1.0             & 84.61          & 76.72                                     & 81.80                             & 62.11                                \\ \hline
0.0             & 1.0             & 83.29          & 76.07                                     & 80.96                             & 60.18                                \\ \hline
1.0             & 0.5             & 81.54          & 73.72                                     & 78.04                             & 56.92                                \\ \hline
1.0             & 0.0             & 76.19          & 70.64                                     & 76.12                             & 52.15                                \\ \hline
0.0             & 0.0             & 74.14          & 69.38                                     & 75.93                             & 50.65                                \\ \hline
\end{tabular}
\vspace{-0.21cm}
\end{table}

$\lambda$ in Eq. (\ref{eq:g-loss}) serves as a loss adjustment factor, where $\lambda_1$ regulates the BN loss and $\lambda_2$ controls the AD loss. To evaluate their impact, we conducted experiments with varying $\lambda_1$ and $\lambda_2$, with results summarized in Table \ref{table:lamda}. The findings indicate that when $\lambda_1 = \lambda_2 = 0$, both BN and AD losses are inactive, leading to a significant decline in FedHydra's performance. Notably, variations in AD loss ($\lambda_2$) have a more pronounced effect compared to BN loss ($\lambda_1$). Overall, both losses contribute significantly to enhancing the generator's performance within the FedHydra framework.

\subsection{Visualization of Synthetic Data}
\begin{figure}[]
\centering
\setlength{\abovecaptionskip}{0.3cm}
\includegraphics[width=0.95\linewidth,scale=1.0]{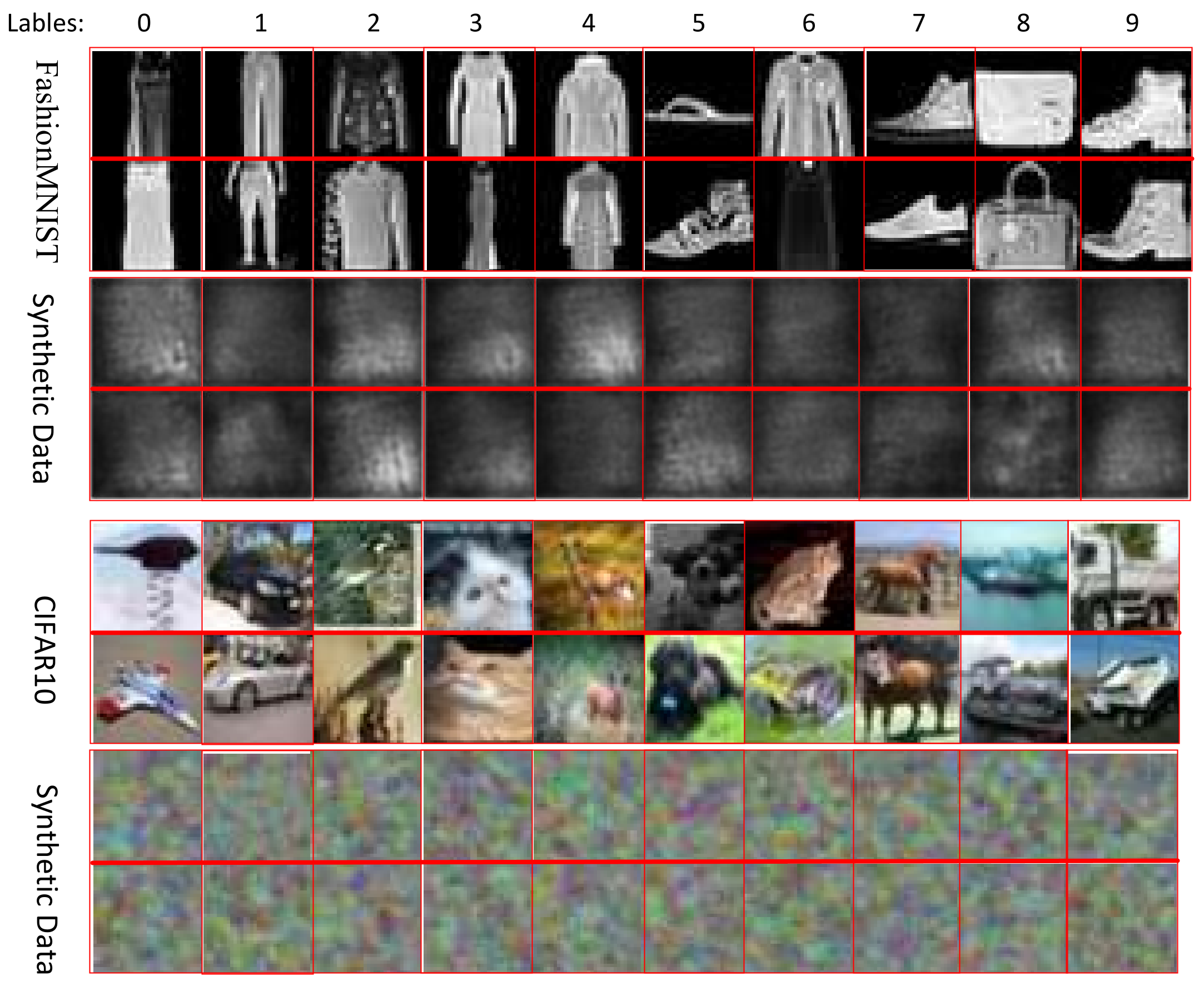}
\caption{Visualization of synthetic data generated by the generator about FashionMNIST and CIFAR10.}
\label{fig:visualization-synthetic-data}
\end{figure}

To assess the quality and privacy of the generated data, we compare synthetic samples with the original training data from FashionMNIST and CIFAR-10 in Figure~\ref{fig:visualization-synthetic-data}. The first and third rows show real samples, while the second and fourth rows present synthetic ones. The synthetic data lack any visually interpretable textures or semantic structures, making them clearly distinct from real data and reducing privacy risks. Despite this, they effectively support model training and enable our method to outperform baseline approaches. These results demonstrate that synthetic data can be both privacy-preserving and performance-enhancing.

\section{Conclusion}

In this work, we present FedHydra, a data-free OSFL framework designed to tackle both data and model heterogeneity. Unlike existing OSFL approaches that address only one aspect of heterogeneity, FedHydra employs a two-stage learning strategy: \textbf{MS} to evaluate client models’ classification strengths and \textbf{HASA} to refine aggregation via a data-free distillation framework. Experimental results across multiple benchmarks demonstrate that FedHydra consistently surpasses SOTA OSFL methods in both homogeneous and heterogeneous environments. FedHydra provides a practical and scalable solution for OSFL. Future research will focus on extending its adaptability to more complex heterogeneity scenarios and optimizing computational efficiency.








\begin{acks}
We sincerely thank all the anonymous reviewers for providing valuable feedback. This work is supported by the NOVA-FRQNT-NSERC grant (https://doi.org/10.69777/328677), Canada Research Chair (Tier 2) in Machine Learning for Genomics and Healthcare (CRC-2021-00547), Natural Sciences and Engineering Research Council (NSERC) Discovery Grant (RGPIN-2016-05174), and Australian Research Council (ARC) Discovery Project Grant (DP230102828).
\end{acks}

\bibliographystyle{ACM-Reference-Format}
\balance
\bibliography{kdd}

\newpage

\appendix

\begin{figure}[H]
\centering
\setlength{\abovecaptionskip}{0.3cm}
\includegraphics[width=\linewidth,scale=1.0]{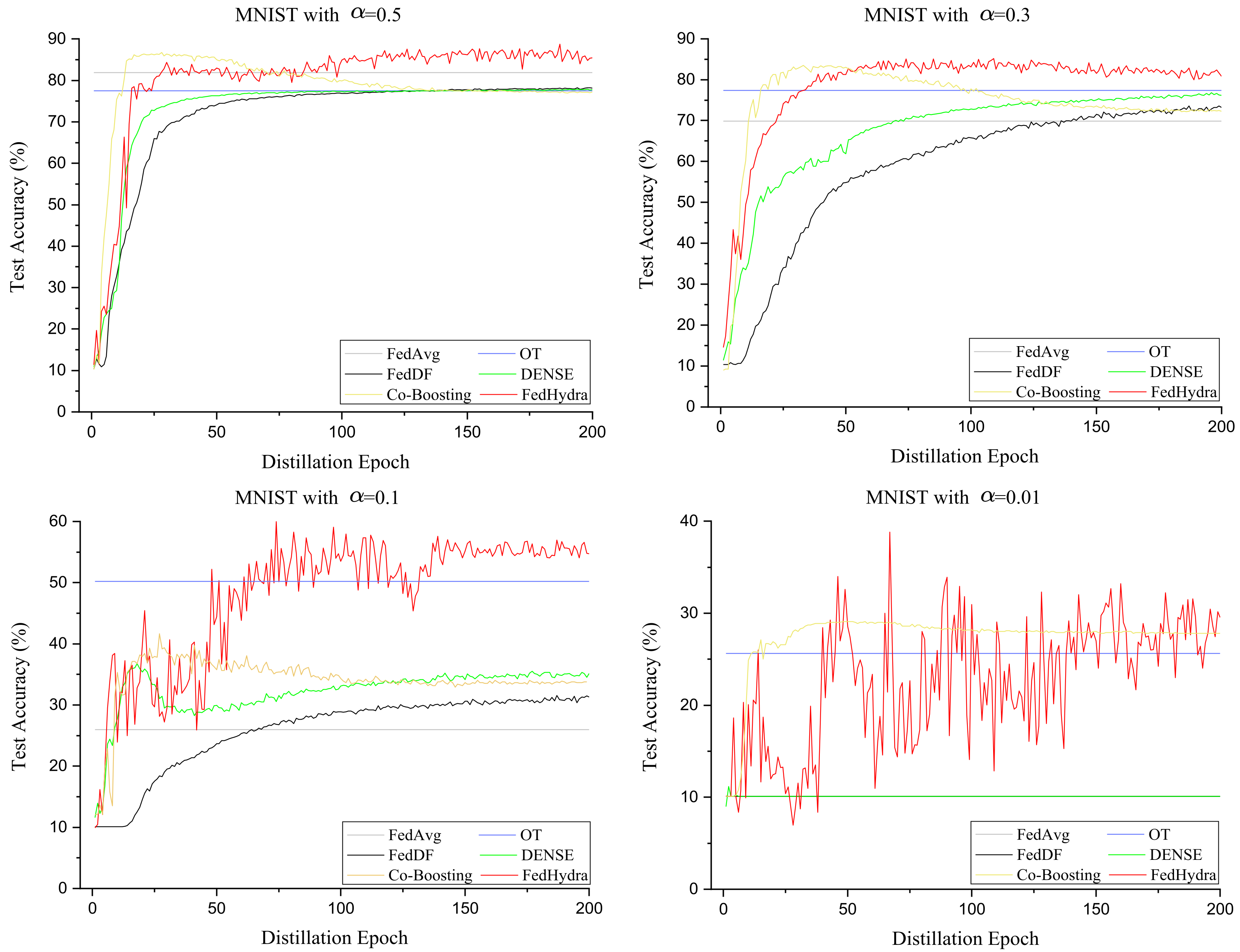}
\caption{Test accuracy versus distillation epoch over MNIST with $\alpha=\{0.5,0.3,0.1,0.01\}$.}
\label{fig:po-mnist-acc}
\end{figure}

\begin{figure}[H]
\centering
\setlength{\abovecaptionskip}{0.1cm}
\includegraphics[width=\linewidth,scale=1.0]{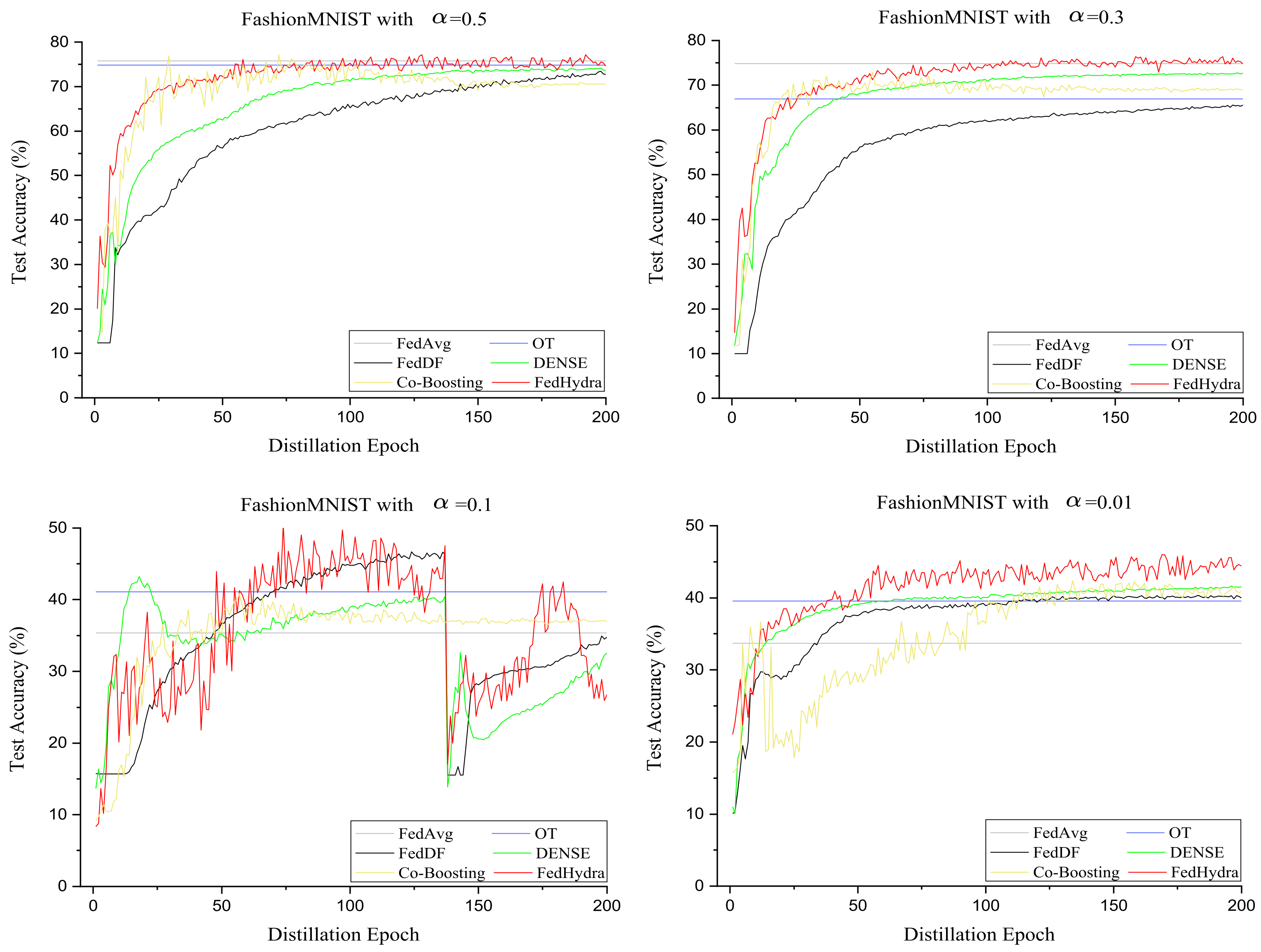}
\caption{Test accuracy versus distillation epoch over FashionMNIST with $\alpha=\{0.5,0.3,0.1,0.01\}$.}
\label{fig:po-fmnist-acc}
\end{figure}

\begin{figure}[H]
\centering
\setlength{\abovecaptionskip}{0.3cm}
\includegraphics[width=\linewidth,scale=1.0]{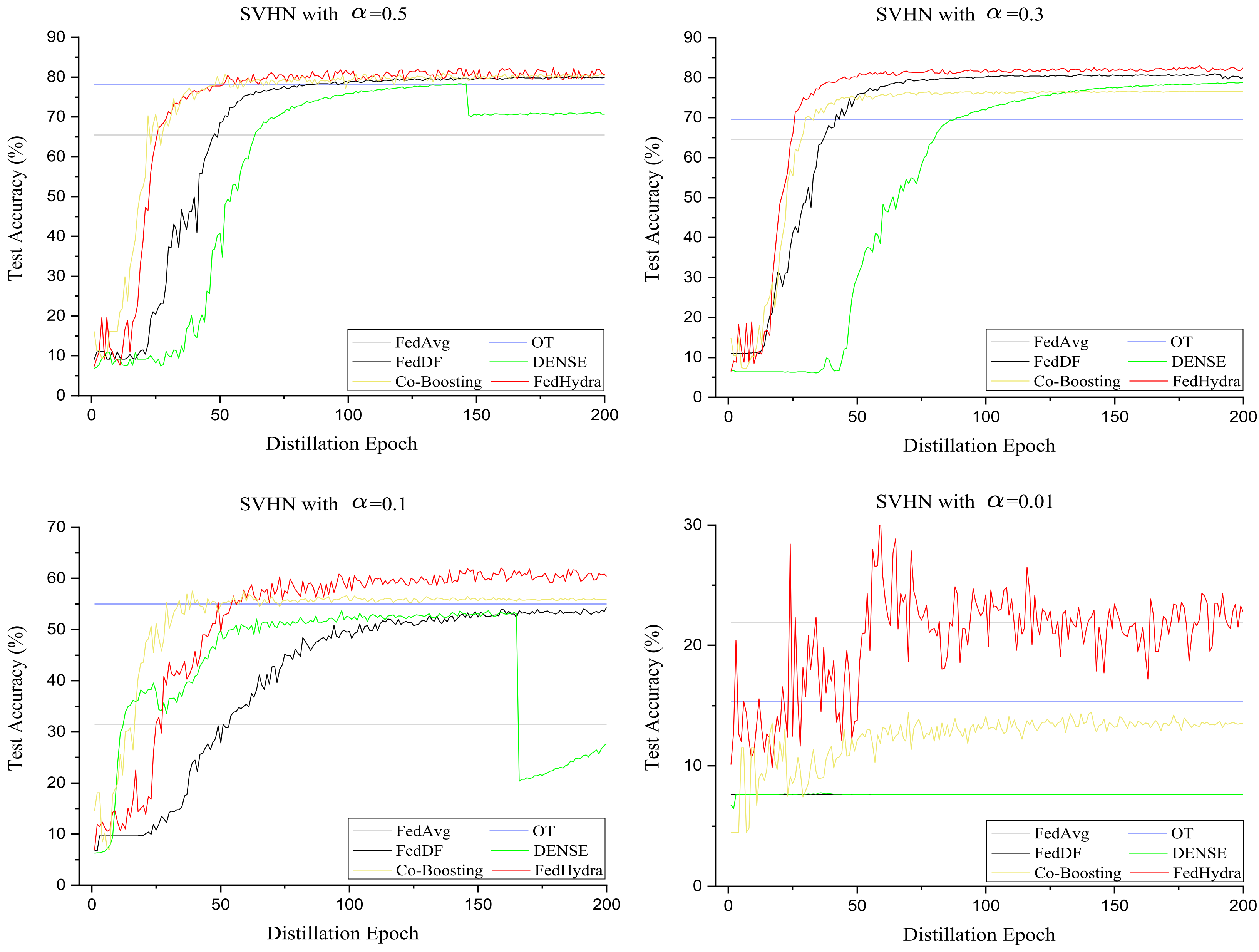}
\caption{Test accuracy versus distillation epoch over SVHN with $\alpha=\{0.5,0.3,0.1,0.01\}$.}
\label{fig:po-svhn-acc}
\end{figure}

\section{Additional Performance Overview Results on MNIST, FashionMNIST, and SVHN} \label{appendix_sec:po}

Figure~\ref{fig:po-mnist-acc}, Figure~\ref{fig:po-fmnist-acc}, and Figure~\ref{fig:po-svhn-acc}  illustrate the test accuracy versus distillation epochs on MNIST, FashionMNIST, and SVHN, respectively, under varying degrees of data heterogeneity ($\alpha = \{0.5, 0.3, 0.1, 0.01\}$). For MNIST, FedHydra surpasses all baseline methods after around 15 epochs and maintains a peak accuracy of approximately 35.00\% near epoch 60 under extreme heterogeneity ($\alpha = 0.01$), while others plateau around 10\%. On SVHN, FedHydra overtakes baselines after 50 epochs and demonstrates a consistent upward trend, peaking around epoch 60 when $\alpha = 0.01$. Similarly, on FashionMNIST, FedHydra shows strong performance across all levels of heterogeneity. It maintains a steady increase in accuracy and achieves around 45\% at $\alpha = 0.01$, clearly outperforming Co-Boosting and other baseline methods that converge to lower values. These results collectively highlight FedHydra's robustness and superior generalization under non-IID conditions.

\section{Extended Analysis}

\subsection{Impact of Client Numbers}
\label{sec:C.1}

Furthermore, we also test the impact of client numbers on CIFAR10 by varying the numbers of clients $K = \{5, 10, 20, 50, 100\}$, as shown in Table \ref{table:impact-cn-cifar10}. All the methods decrease when the client number increases. However, FedHydra has less impact than other methods,  FedHydra achieves $47.40$\% when $K = 100$ and outperforms the second-best DENSE with only $36.25$\%. The proposed SA algorithm enables FedHydra to scale effectively, exhibiting minimal sensitivity to the increasing number of clients.

\begin{table}[H]
\setlength{\abovecaptionskip}{5pt}
\setlength{\tabcolsep}{2pt} 
\fontsize{8}{10}\selectfont
\caption{Top-1 test accuracy achieved by FedAvg, OT, FedDF, DENSE, Co-Boosting, and FedHydra over CIFAR10 with different client numbers $K=\{5,10,20,50,100\}$.}
\label{table:impact-cn-cifar10}
\begin{tabular}{c|cccccc}
\hline
Client $K$ & FedAvg  & OT  & FedDF  & DENSE & Co-Boosting & FedHydra  \\ \hline
5   & 38.26 & 52.88 & 54.81  & 58.48 & 62.88 & \textbf{63.15}  \\ \hline
10  & 41.71 & 47.86 & 50.17  & 55.43 & 61.56  & \textbf{61.58}  \\ \hline
20  & 33.47 & 44.26 & 38.70  & 42.50 & 53.49  & \textbf{59.64}  \\ \hline
50  & 31.75 & 40.37 & 37.85  & 39.09 & 45.21  & \textbf{52.66}  \\ \hline
100 & 27.23 & 37.18 & 33.90  & 36.25 & 38.96  & \textbf{47.40}  \\ \hline
\end{tabular}
\end{table}

\subsection{Performance Under Multiple Global Training Rounds}
\label{sec:C.2}

We extend our experiments to multiple global rounds. Specifically, we fix the local epoch $E=50$ and $\alpha = 0.1$ to evaluate the performance of diverse methods on CIFAR10 with varying client distributions. Table \ref{table:mr-svhn-a0.1} summarizes the relevant evaluation results.  The results in Table \ref{table:mr-svhn-a0.1} show that FedHydra outperforms the other methods with different global training rounds from $T = \{1, 2, 3, 4, 5\}$. FedHydra steadily increases and reaches a peak when $T = 3$, which outperforms DENSE by $1.86$\%.  

\begin{table}[]
\setlength{\abovecaptionskip}{5pt}
\setlength{\tabcolsep}{2pt} 
\fontsize{8}{10}\selectfont
\caption{Top-1 test accuracy achieved by FedAvg, OT, FedDF, DENSE, Co-Boosting,  and FedHydra over SVHN with $\alpha=0.1$ under multiple global rounds.}
\label{table:mr-svhn-a0.1}
\begin{tabular}{c|cccccc}
\hline
\begin{tabular}[c]{@{}c@{}}Global \\ round\end{tabular} & FedAvg  & OT  & FedDF  & DENSE & Co-Boosting & FedHydra  \\ \hline
$T$=1  & 28.31 & 41.85 & 48.13  & 48.22 & 46.87 & \textbf{53.27} \\ \hline
$T$=2  & 34.55 & 44.32 & 56.62  & 54.67 & 57.24 & \textbf{59.98} \\ \hline
$T$=3  & 44.60 & 52.29 & 65.61  & 68.44 & 65.14 & \textbf{70.30} \\ \hline
$T$=4  & 57.05 & 53.96 & 63.62  & 66.83 & 63.59 & \textbf{68.58} \\ \hline
$T$=5  & 61.59 & 64.01 & 66.66  & 64.57 & 64.07 & \textbf{70.39} \\ \hline
\end{tabular}
\end{table}

\section{Additional Experimental Results on SVHN under OSFL Model Heterogeneity}
\label{sec:D}


Table~\ref{table:hfl-svhn} reports top-1 test accuracy on SVHN under both data and model heterogeneity, using five heterogeneous client models and ResNet18 as the server. As $\alpha$ decreases, all methods suffer performance drops, but FedHydra consistently outperforms baselines. At $\alpha = 0.01$, it achieves 54.33\%, notably higher than FedDF (43.82\%) and DENSE (42.39\%). Even under moderate heterogeneity ($\alpha = 0.3, 0.5$), FedHydra maintains the highest accuracy, confirming its robustness across heterogeneous settings.

\begin{table}[H]
\setlength{\abovecaptionskip}{5pt}
\setlength{\tabcolsep}{1.5pt}
\fontsize{7}{10}\selectfont
\caption{Top-1 test accuracy achieved by FedDF, DENSE, Co-Boosting, and FedHydra over SVHN with personalized client models. M1 to M5 represent different neural network architectures: GoogleNet, ResNet18, two custom CNN models, and LeNet, respectively.}
\label{table:hfl-svhn}
\begin{tabular}{c|ccccc|cccc}
\hline
\multirow{2}{*}{\begin{tabular}[c]{@{}c@{}} $\alpha$ \end{tabular}} & \multicolumn{5}{c|}{Personalized Client Model}                                        & \multicolumn{3}{c}{Server (ResNet18)} \\ \cline{2-10} 
                                                                                & \multicolumn{1}{c}{M1} & \multicolumn{1}{c}{M2} & M3  & M4  & M5 & FedDF        & DENSE  &
                                                                Co-Boosting      & FedHydra              \\ \hline
$\alpha$=0.01                                                                          & 24.17                         & 7.59                         & 29.81 & 34.76 & 19.52 & 43.82        & 42.39  & 45.72        & \textbf{54.33}        \\ \hline
$\alpha$=0.1                                                                           & 42.28                         & 14.84                        & 39.64 & 34.80 & 31.17 & 37.43        & 36.00 & 44.25        & \textbf{47.45}        \\ \hline
$\alpha$=0.3                                                                           & 66.78                         & 65.62                        & 38.22 & 58.73 & 64.19 & 77.22        & 75.75  & 73.21        & \textbf{77.52}        \\ \hline
$\alpha$=0.5                                                                           & 72.54                         & 70.72                        & 39.74 & 69.24 & 64.56 & 69.75        & 64.67 & 69.93        & \textbf{71.29}        \\ \hline
\end{tabular}
\end{table}


Table~\ref{table:hfl-2CC} shows top-1 test accuracy on SVHN and CIFAR10 under the challenging 2c/c setting in heterogeneous OSFL. Across both datasets, FedHydra consistently outperforms other methods. On SVHN, it achieves 31.22\%, surpassing Co-Boosting (27.34\%), FedDF (24.72\%), and DENSE (20.42\%). Similarly, on CIFAR10, FedHydra reaches 38.57\%, outperforming Co-Boosting (33.54\%), FedDF (31.05\%), and DENSE (26.18\%). These results highlight FedHydra's superior robustness under extreme heterogeneity.

\begin{table}[H]
\setlength{\abovecaptionskip}{5pt}
\setlength{\tabcolsep}{1.5pt}
\fontsize{7}{10}\selectfont
\caption{Top-1 test accuracy achieved by FedDF, DENSE, Co-Boosting, and FedHydra over SVHN and CIFAR10 with 2c/c in heterogeneous OSFL. M1 to M5 represent different neural network architectures: GoogleNet, ResNet18, two custom CNN models, and LeNet, respectively.}
\label{table:hfl-2CC}
\begin{tabular}{c|ccccc|cccc}
\hline
\multirow{2}{*}{\begin{tabular}[c]{@{}c@{}}Dataset\\ with 2c/c\end{tabular}} & \multicolumn{5}{c|}{Personalized Client Model}                                        & \multicolumn{3}{c}{Server (ResNet18)} \\ \cline{2-10} 
                                                                             & \multicolumn{1}{c}{M1} & \multicolumn{1}{c}{M2} & M3  & M4  & M5 & FedDF    & DENSE                                                                   &    Co-Boosting     & FedHydra         \\ \hline
SVHN                                                                         & 25.78                         & 26.30                        & 18.48 & 15.13 & 11.85 & 24.72    & 20.42  & 27.34 & \textbf{31.22}   \\ \hline
CIFAR10                                                                      & 19.43                         & 18.10                        & 18.69 & 19.62 & 19.00 & 31.05    & 26.18   & 33.54 & \textbf{38.57}   \\ \hline
\end{tabular}
\end{table}

\section{Additional Experimental Results Under Multiple Global Rounds Across Diverse $\alpha$ Values}
\label{sec:E}

Tables \ref{table:mr-cifar10-a0.5}, \ref{table:mr-cifar10-a0.3}, and \ref{table:mr-cifar10-a0.01} provide the relevant test results with $\alpha=\{0.5, 0.3, 0.01\}$  under five global rounds on CIFAR10.


\begin{table}[H]
\setlength{\abovecaptionskip}{5pt}
\setlength{\tabcolsep}{2pt} 
\fontsize{8}{10}\selectfont
\caption{Top-1 test accuracy achieved by FedAvg, OT, FedDF, DENSE, Co-Boosting, and FedHydra over CIFAR10 with $\alpha=0.5$ under multiple global rounds.}
\label{table:mr-cifar10-a0.5}
\begin{tabular}{c|cccccc}
\hline
\begin{tabular}[c]{@{}c@{}}Global \\ round\end{tabular} & FedAvg  & OT  & FedDF  & DENSE & Co-Boosting & FedHydra  \\ \hline
$T$=1  & 38.26 & 47.56 & 48.14 & 50.23 & 50.68 & \textbf{53.48} \\ \hline
$T$=2  & 41.71 & 48.93 & 49.26 & 52.58 & 55.21 & \textbf{57.09} \\ \hline
$T$=3  & 45.42 & 50.33 & 48.71 & 55.39 & 57.20 & \textbf{62.75} \\ \hline
$T$=4  & 49.81 & 53.61 & 48.86 & 53.81 & 57.66 & \textbf{61.83} \\ \hline
$T$=5  & 51.75 & 53.80 & 49.70 & 53.49 & 58.72 & \textbf{64.71} \\ \hline
\end{tabular}
\end{table}


\begin{table}[H]
\setlength{\abovecaptionskip}{5pt}
\setlength{\tabcolsep}{2pt} 
\fontsize{8}{10}\selectfont
\caption{Top-1 test accuracy achieved by FedAvg, OT, FedDF, DENSE, Co-Boosting,  and FedHydra over CIFAR10 with $\alpha=0.3$ under multiple global rounds.}
\label{table:mr-cifar10-a0.3}
\begin{tabular}{c|cccccc}
\hline
\begin{tabular}[c]{@{}c@{}}Global \\ round\end{tabular} & FedAvg  & OT  & FedDF  & DENSE & Co-Boosting & FedHydra  \\ \hline
$T$=1  & 33.72 & 36.56 & 37.56 & 41.54 & 48.01 & \textbf{50.01} \\ \hline
$T$=2  & 37.56 & 39.62 & 38.94 & 46.44 & 52.30 & \textbf{57.00} \\ \hline
$T$=3  & 36.04 & 38.24 & 40.73 & 47.10 & 56.06 & \textbf{58.18} \\ \hline
$T$=4  & 41.73 & 43.00 & 41.10 & 46.20 & 55.21 & \textbf{58.43} \\ \hline
$T$=5  & 44.26 & 46.85 & 41.03 & 48.55 & 51.34 & \textbf{55.97} \\ \hline
\end{tabular}
\end{table}


Table~\ref{table:mr-cifar10-a0.5} shows that FedHydra consistently outperforms and remains more stable than other methods across different global rounds when $\alpha = 0.5$. Its accuracy on CIFAR-10 surpasses the second-best method, DENSE, by over 10\% at the 5\textsuperscript{th} round. Table~\ref{table:mr-cifar10-a0.3} further demonstrate FedHydra’s clear advantage under moderate heterogeneity ($\alpha = 0.3$), reaching 86.72\% on SVHN and 58.43\% on CIFAR-10. Under extreme heterogeneity ($\alpha = 0.01$), as shown in Table~\ref{table:mr-cifar10-a0.3}, FedHydra maintains top performance across all rounds, peaking at 39.58\% on SVHN and 39.76\% on CIFAR-10, significantly ahead of all baselines. These results highlight FedHydra’s superior robustness and adaptability in heterogeneous FL.


Table~\ref{table:mr-cifar10-a0.01} presents the top-1 test accuracy on CIFAR-10 under extreme data heterogeneity ($\alpha = 0.01$) across different global training rounds. FedHydra consistently outperforms all baselines at every round, achieving peak accuracy of 39.76\% at $T=3$ and maintaining strong performance in later rounds. In contrast, FedDF performs poorly throughout, while FedAvg and OT show minor fluctuations and gradual improvements. Co-Boosting performs competitively but still lags behind FedHydra. These results highlight FedHydra’s robustness and adaptability under severe non-IID conditions.

\begin{table}[H]
\setlength{\abovecaptionskip}{5pt}
\setlength{\tabcolsep}{2pt} 
\fontsize{8}{10}\selectfont
\caption{Top-1 test accuracy achieved by FedAvg, OT, FedDF, DENSE, Co-Boosting,  and FedHydra over CIFAR10 with $\alpha=0.01$ under multiple global rounds.}
\label{table:mr-cifar10-a0.01}
\begin{tabular}{c|cccccc}
\hline
\begin{tabular}[c]{@{}c@{}}Global \\ round\end{tabular} & FedAvg  & OT  & FedDF  & DENSE & Co-Boosting & FedHydra  \\ \hline
$T$=1  & 10.00 & 16.12  & 14.39 & 23.84 & 26.60 & \textbf{30.13} \\ \hline
$T$=2  & 17.77 & 20.33  & 14.89 & 24.84 & 29.88 & \textbf{37.67} \\ \hline
$T$=3  & 26.02 & 25.98  & 16.42 & 27.24 & 33.68 & \textbf{39.76} \\ \hline
$T$=4  & 25.57 & 29.50  & 16.79 & 29.16 & 34.07 & \textbf{38.17} \\ \hline
$T$=5  & 27.68 & 26.38  & 17.59 & 26.04 & 35.12 & \textbf{38.30} \\ \hline
\end{tabular}
\end{table}

\end{document}